\documentclass{sig-alternate}
\usepackage{listings}
\usepackage{graphicx}
\let\chapter\section
\usepackage[lined,commentsnumbered,linesnumbered,noend]{algorithm2e}
\usepackage{subcaption}
\usepackage[labelfont=bf]{caption}
\usepackage{balance}  

\usepackage[usenames,dvipsnames]{xcolor}

\usepackage{amsfonts}
\usepackage{amssymb}
\usepackage{amsmath}
\usepackage{latexsym}
\usepackage{verbatim}
\usepackage{url}
\usepackage{color}
\usepackage{calc}
\usepackage{array}

\usepackage{textcomp}
\usepackage{paralist}
\usepackage{comment}
\usepackage{pifont,array,amsmath,multirow,latexsym,tabularx,amssymb}
\usepackage[numbers,sort]{natbib}

\usepackage{hyperref}
\usepackage{url}

\usepackage[usenames,dvipsnames]{xcolor}

\usepackage{enumitem}

\usepackage{graphicx}
\usepackage{xspace}
\usepackage{amsmath}
\usepackage{amssymb}
\usepackage{color}

\usepackage{float}
\usepackage{listings}
\usepackage{graphicx}
\usepackage{float}

\newtheorem{theorem}{Theorem}

\usepackage{amsfonts}

\newcommand{\inv}[0]{\vspace*{-0.2cm}}
\newcommand{\sinv}[0]{\vspace*{-0.1cm}}

\newcommand*{\bmfontA}{\fontfamily{fvs}\selectfont}
\newcommand*{\bmfontB}{\fontfamily{fi4}\selectfont}
\newcommand*{\bmfontC}{\fontfamily{lmss}\selectfont}
\newcommand*{\bmfontD}{\fontfamily{put}\selectfont}
\newcommand*{\bmfontE}{\fontfamily{ppl}\selectfont}
\newcommand*{\bmfontF}{\fontfamily{cmr}\selectfont}
\DeclareTextFontCommand{\bmfont}{\bmfontB}
\DeclareTextFontCommand{\bmfontt}{\bmfontA}
\DeclareTextFontCommand{\bmfonttt}{\bmfontC}
\DeclareTextFontCommand{\bmfontttt}{\bmfontD}
\DeclareTextFontCommand{\bmfonttttt}{\bmfontE}
\DeclareTextFontCommand{\bmfontttttt}{\bmfontF}
\DeclareTextFontCommand{\bmfonttttttt}{\bmfontG}

\newcommand{\vprofiler}[0]{\bmfont{VProfiler}}
\newcommand{\vats}[0]{\bmfont{VATS}}

\newcommand{\ignore}[1]{}

\newcommand{\ph}[1]{\vspace{2mm} \noindent \textbf{#1} ---}
\newcommand{\mph}[1]{\vspace{1mm} \noindent \textbf{$\bullet~$ #1:}}

\newcommand{\tofix}[1]{\textcolor{red}{#1}}

\newcommand{\cancut}[1]{\textcolor{black}{#1}}
\newcommand{\tempcut}[1]{#1}

\definecolor{mypink3}{cmyk}{0, 0.7808, 0.4429, 0.1412}
\newcommand{\jiamin}[1]{\textcolor{mypink3}{Jiamin: #1}}

\begin{document}
\setcopyright{acmcopyright}
\conferenceinfo{SIGMOD '16}{June 26--July 1, 2016, San Fransisco, CA, USA}
\CopyrightYear{2016} 
\crdata{0-89791-88-6/97/05}


\title{Identifying the Major Sources of Variance in Transaction Latencies:
Towards More Predictable Databases}



%
%
%
%

\numberofauthors{1} 

\author{
%
%
\alignauthor
Jiamin Huang~~~~~~~~Barzan Mozafari~~~~~~~~Grant Schoenebeck~~~~~~~~Thomas Wenisch\\
\vspace{0.15cm}
University of Michigan, Ann Arbor\\
\vspace{0.1cm}
{\large \{jiamin, mozafari, schoeneb, twenisch\}@umich.edu}
}

\maketitle


\begin{abstract}
Decades of research have sought to improve transaction processing performance and scalability in database management systems (DBMSs).  
However, significantly less attention has been dedicated to the predictability of performance: how often individual transactions exhibit execution latency far from the mean?
Performance predictability is vital when transaction processing lies on the critical path of a complex enterprise software or an interactive web service, 
as well as in emerging database-as-a-service markets where customers contract for guaranteed levels of performance.
In this paper, we take several steps towards achieving more predictable database systems.  
First, we propose a profiling framework called \vprofiler\ that, given the source code of a DBMS, is able to identify the dominant sources of variance in transaction latency. \vprofiler\ automatically instruments the DBMS source code to deconstruct the overall variance of transaction latencies
 into variances and covariances of the execution time of individual functions, which in turn provide insight into the root causes of variance. 
Second, we use \vprofiler\ to analyze MySQL and Postgres---two of the most popular and complex open-source
database systems.
Our case studies reveal that the primary causes of variance in MySQL and Postgres are 
       lock scheduling  and centralized logging, respectively.
Finally, based on \vprofiler's findings, we further focus on remedying the performance variance of MySQL  by 
(1) proposing a new lock scheduling algorithm, called Variance-Aware Transaction Scheduling (\vats), 
(2) enhancing the buffer pool replacement policy, and (3) identifying tuning parameters that can reduce variance significantly. 
 Our experimental results show that our schemes reduce overall transaction latency variance by 37\% on average (and up to 64\%) without compromising throughput or mean latency. 
 \end{abstract}


\section{Introduction}
\label{sec:intro}

Transactional databases are a key component of almost every enterprise software system,
where mission-critical applications rely on database management systems to store and manipulate  data efficiently. 
Consequently, a significant portion of database research on transactions has focused on reducing latency and 
increasing scalability and throughput, for example,  through
 concurrency control, query optimization techniques, indexing, 
 caching, and other sophisticated ideas. 
These strategies, however, have been vetted primarily in terms of
 their effect on the \emph{average performance} of the database, such as its  
 mean transaction latency or throughput. 
  In other words, the focus has been on understanding average performance and running more and/or faster transactions \emph{overall}.

While peak transaction processing throughput is clearly an important metric, the \emph{predictability} 
of performance---the disparity between average and high-percentile tail latencies---has emerged 
as an
 equally important metric in many situations. However, performance predictability
  has often been ignored in traditional efforts to improve throughput
and mean latency.
In fact, some widely adopted optimization strategies (e.g., asynchronous logging and group commit~\cite{helland1989group,strom1985optimistic}) deliberately 
improve throughput at the expense of penalizing latency for \emph{some} transactions. 
While the overall  breakdown of mean transaction latency
 in terms of various database components has been carefully studied~\cite{looking-glass},  
 an analogous study to identify the sources of latency variance has not been attempted before.

\begin{figure}[t!]
\centering
\includegraphics[scale=0.1]{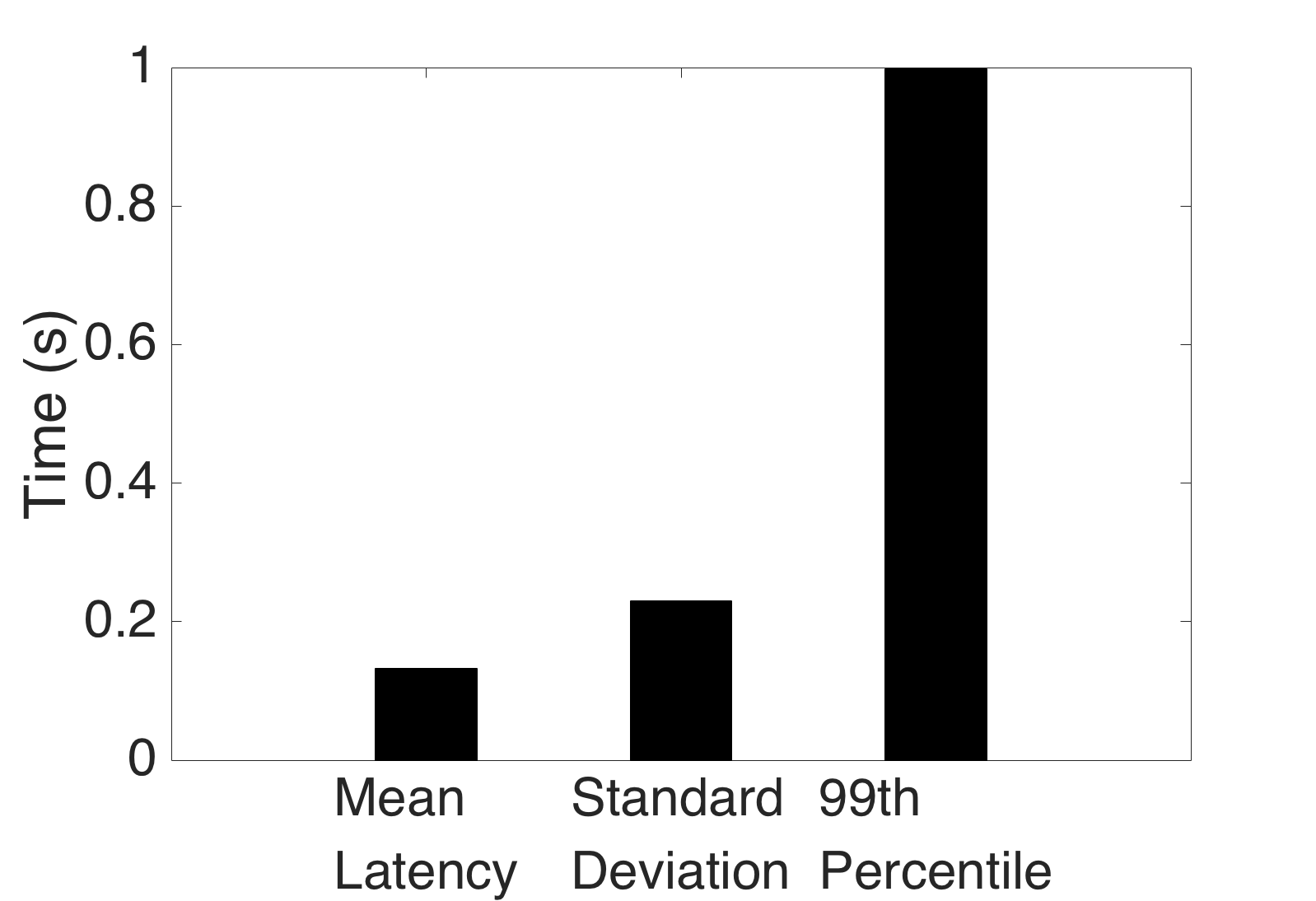}
\caption{99th percentile latency, standard deviation and average latency in MySQL.}
\inv\sinv
\label{fig:mysql-var}
\end{figure}

\begin{figure}[t]

\centering
\includegraphics[scale=0.1]{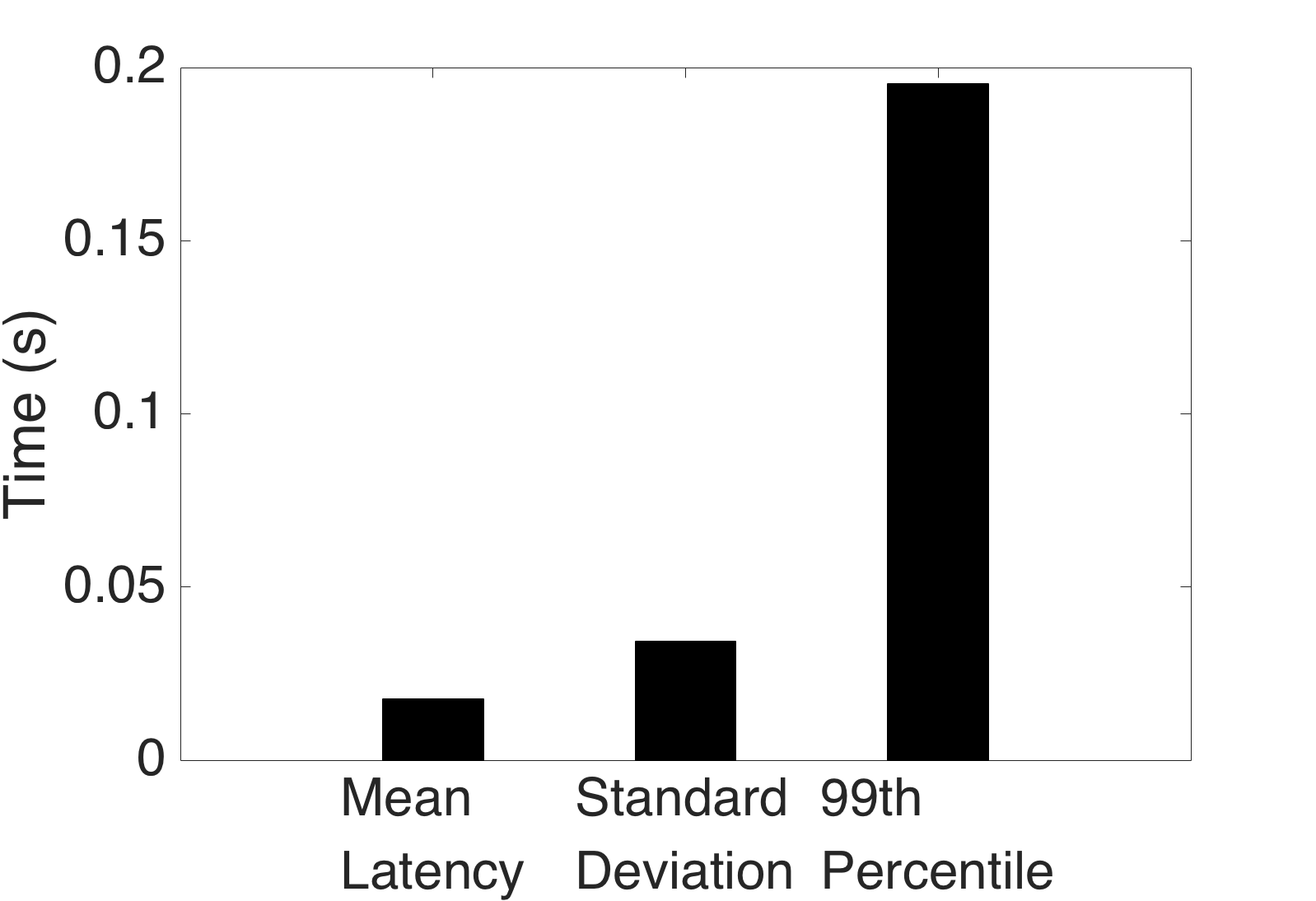}
\caption{99th percentile latency, standard deviation and average latency in Postgres.}
\inv\sinv
\label{fig:post-var}
\end{figure}

\begin{figure}[t]
\centering
\includegraphics[scale=0.1]{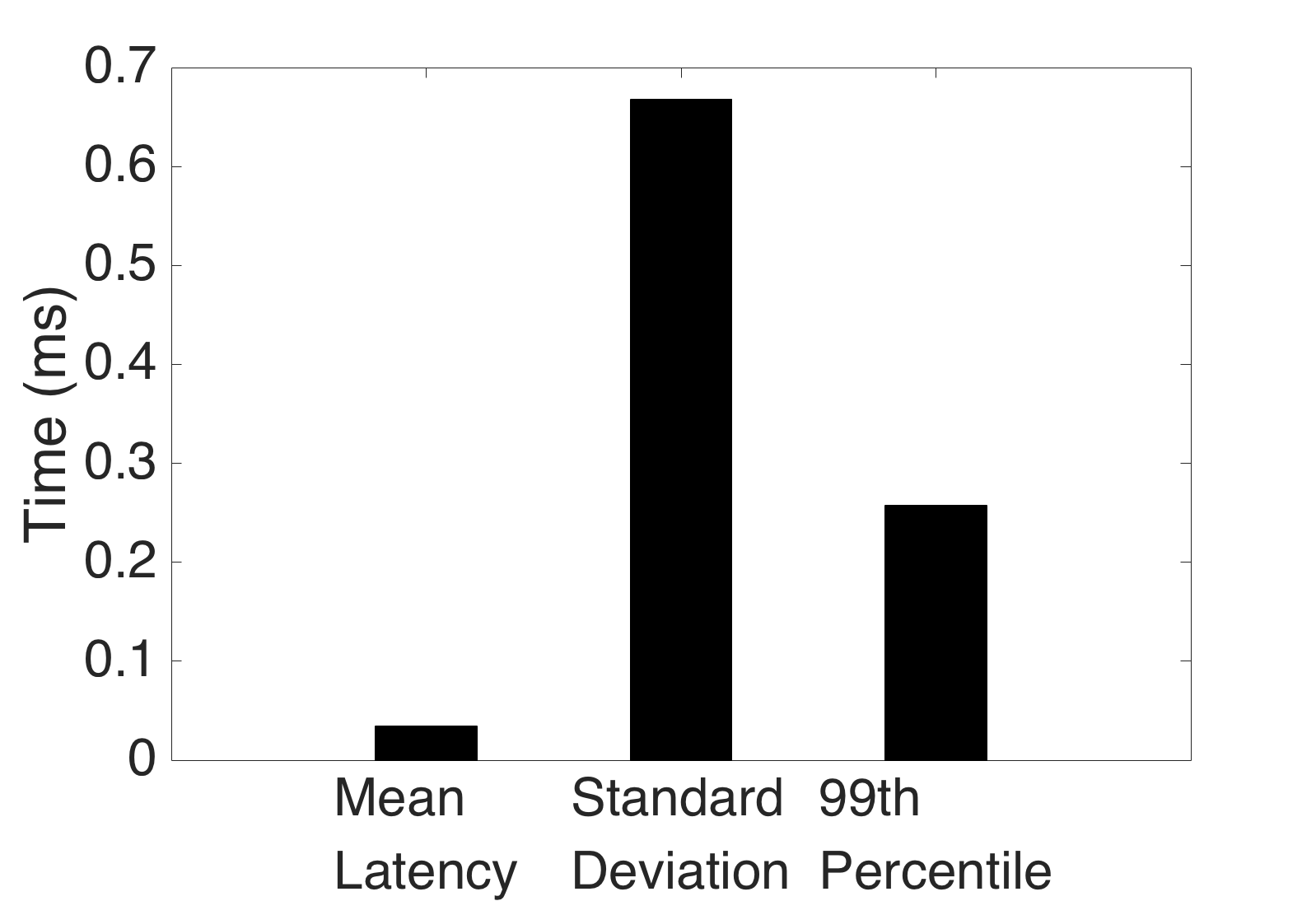}
\caption{99th percentile latency, standard deviation and average latency in VoltDB.}
\inv\sinv
\label{fig:voltdb-var}
\end{figure}

At the fine time scale of individual transactions, database performance is astonishingly unpredictable, with a large gap between mean and high percentile transaction latency.
Figure~\ref{fig:mysql-var} illustrates the magnitude of transaction latency variance in MySQL.
The figure shows the mean, standard deviation, and 99th percentile latencies 
observed in the TPC-C online transaction processing benchmark (see Section~\ref{sec:mysql} for methodology
details).
The transaction latency standard deviation is nearly twice the mean, and the 99th percentile latency is
an order of magnitude larger. 
This wide performance variability is not limited to MySQL, and is exhibited by most database systems 
on the market, e.g., similar ratios are observed in Postgres and VoltDB, as is shown in figure \ref{fig:post-var} and \ref{fig:voltdb-var}.

\ignore{Note that certain amount of performance disparity across different  types of transactions is inevitable. For example, 
a New Order transaction in TPC-C involves significantly more \emph{work} than a Stock Level transaction. 
However, we still observe a similar ratio between high percentile and mean latency even among transactions performing the same amount of work, 
e.g, two New Order transactions with the same number of items.}


Advancements in hardware and storage devices and new business models for providing database-as-a-service have increased the need to study and optimize latency variance.
First, faster storage, increased hardware parallelism, and better transaction processing schemes
have enabled microsecond latencies and thousands of concurrent transactions~\cite{bailis2015coordination,lahiri2013oracle,thousand-cores}.
As mean performance improves, the impact of performance perturbations (e.g., due to a slow I/O
request) relative to the latency of a transaction grows. 
 Second, database vendors are facing an increasing number of business-oriented clients and applications 
that demand quality of service guarantees (QoS).
Moreover, with the increasing market share of database-as-a-service  (DBaaS) offerings,
cloud providers and users rely on service level agreements (SLAs) for pricing and provisioning, respectively~\cite{sqlvm-cidr,sqlvm-BS,oracle-cloud,google-cloud,mozafari_cidr2013}.
Failing to meet performance objectives, even for a subset of transactions or users, 
can result in financial penalties for the DBaaS provider.
Finally, as DBMSs deliver 
a wide range of complex features to a wide range of applications,
they have (understandably) become one of the most complex breeds of software systems.
Subtle interactions of difficult-to-analyze code paths lead to vexing performance anomalies.

In light of these trends, we believe it is critical to undertake a systematic approach to managing
performance variance in transaction processing systems.
Some of this variance is inherent in the nature of the transactions; some must perform more work
than others.
Nevertheless, our study reveals that dominant sources of variance are often not a function of
work, and rather arise due to scheduling, contention, I/O, and other less predictable effects.
Understanding the major sources of variance in the execution time of transactions 
can provide 
invaluable insight towards designing a new generation of database systems that can deliver
\emph{competitive performance while being much more predictable}.
A predictable database has a myriad of benefits: (i)  
meeting SLAs with fewer resources due to a reduced need for over-provisioning, 
(ii) more accurate cost estimates, and hence, better query scheduling and planning decisions,
and (iii) easier performance tuning and diagnosis.  
Identifying the major sources of variance is the first step to
achieving predictable performance in any database system.

In this work, we propose a tool called \vprofiler\  that automatically instruments the source code of a DBMS to
quantify the dominant sources of variance in transaction latency and help identify
those that are inherent to the amount of work a transaction performs and those that arise due to a performance pathology. 
To minimize the overhead of  collecting fine-grain performance measurements,  
\vprofiler\  runs in multiple iterations, each time instrumenting  
a carefully selected subset of functions invoked during transaction processing. 
  By analyzing these measurements at each iteration, 
 \vprofiler\  deconstructs overall variance of  transaction latencies into variances and covariances 
of the execution time of individual functions, providing insight into the root causes of performance variance.


Through a careful case study of transaction processing in MySQL and Postgres (two popular open-source DBMSs),
	we evaluate \vprofiler's effectiveness. Based on \vprofiler's findings, we also propose several strategies for reducing performance variance 
	in both systems.
		Besides these concrete contributions for MySQL and Postgres, which yield immediate practical benefits,
		we hope that our framework and findings serve as a beginning step towards  a new  
			generation of predictable databases that treat performance variance as a first-class citizen (in addition to peak performance).

\ph{Previous Approaches} There has been some pioneering work on enriching 
query optimizers to 
account for parameter uncertainties (caused by cost or cardinality estimates) when choosing a query plan~\cite{variance-aware}. 
Others have taken the opposite approach by 
 always resorting to table scans for all queries~\cite{constant-qp, least-expected-query,shared-scan2,complaint-paper,qpipe,datapath}, or 
 simply restricting themselves to query plans with a bounded worst-case~\cite{piql}. 
Whereas many of these techniques try to share the scans and joins across multiple queries, they naturally increase the latency of 
individual queries,  and can therefore have a negative impact on average latency.
 As a result, despite their many merits, these techniques are not adopted by any of the major DBMSs, as foregoing low latency to achieve predictability 
 	is typically not a compelling trade-off.

Instead of requiring richer statistics or dismissing traditional query optimizers altogether, 
	in this paper we take a different approach by carefully studying the source code of popular database systems,
	to understand the root causes of performance variance. 
Moreover, we focus our attention on techniques that can reduce tail latency without sacrificing throughput or average latency.
Although building a new DBMS from scratch that is designed for predictability might be a tempting and worthwhile 
endeavor---we believe that understanding the  major sources of latency variance in today's databases offers invaluable insight for guiding the future DBMS design.
	In fact, even in the short term, enhancing the predictability of  DBMSs such as MySQL and Postgres
		is a worthy cause that can impact thousands of DBAs, application developers, and millions of 
		 users interacting with applications that are backed by these popular DBMSs.

\ignore{Blink~\cite{shared-scan2} and Crescando~\cite{complaint-paper} share scans at the storage engine level. QPipe~\cite{qpipe}, DataPath~\cite{datapath}, SharedDB~\cite{shareddb}, and CJoin~\cite{cjoin} add more complex operators and try to avoid repetitive work by sharing it across all queries. Instead of instantiating operators at runtime, most of these systems use a set of always-on operators in order to further reduce execution time and maximize sharing.}

\ph{Contributions}
We make the following contributions:
\begin{enumerate}[leftmargin=0.25cm]

\item We present  \vprofiler, as the first profiling tool that can efficiently and rigorously decompose the variance of the overall execution time of an application
	by automatically instrumenting its source code, and identifying the major functions that contribute the most to the overall variance (Section~\ref{sec:vp}).\footnote{\vprofiler\ is open source: \url{http://github.com/mozafari/vprofiler}} 

\item We use \vprofiler\ to analyze MySQL codebase and find that varying delays due to lock scheduling 
are a dominant source
of  latency variance. \vprofiler\ also finds the LRU policy as another cause  of variance, 
when working sets exceed buffer pool capacity (Section~\ref{sec:mysql}).

\item We further evaluate \vprofiler\ by analyzing Postgres codebase and finding that variance in the delay to flush redo logs  accounts for 70\% of overall latency variance.  Unlike MySQL's delta logging, Postgres logs modified rows in their 
entirety. This, combined with centralized 
  logging, leads to latency variance (Section~\ref{sec:psql}).

\item While most DBMSs grant locks on a first-come-first-served basis, we propose a \emph{variance-aware transaction scheduling} (\vats) algorithm, 
	which lends itself to an adaptive lock manager.
	By minimizing the $L_p$ norm, \vats\ simultaneously reduces mean, variance and high percentiles of transaction latencies.  We prove that, in the absence of any prior knowledge on transactions' remaining times, \vats\ is the optimal strategy.

\item We propose other variance reduction strategies specific to the sources of variance in MySQL and Postgres, including 
 a \emph{lazy LRU update} policy that significantly reduces
	  contention on the buffer pool manager. As a practical guideline for database administrators, we also 
	  suggest \emph{variance-aware tuning} strategies to further reduce variance. (Section~\ref{sec:predict:specific})

\item We provide extensive  experiments across five different benchmarks with varying complexity,
	confirming that our techniques reduce mean, variance, and 99th percentile latencies on average by 
	26\%, 37\%, and 36.8\%, respectively (and up to 59.7\%, 64\%, and 64.4\%, resp.) 
	 without compromising throughput (Section~\ref{sec:experiments}). 

\end{enumerate}

We introduce \vprofiler\ in Section~\ref{sec:vp}.
In Sections~\ref{sec:mysql} and~\ref{sec:psql}, we present our case studies of transaction latency variance in MySQL and Postgres, respectively.
We describe our general and DBMS-specific variance-reduction strategies in Sections~\ref{sec:vats} and~\ref{sec:predict:specific}, respectively. 
We present experimental results in Section~\ref{sec:experiments}, and discuss the related work in Section~\ref{sec:related}.




\section{VProfiler}
\label{sec:vp}
With the complexity of modern software,  
 there are many possible causes of latency variance, such as I/O operations, locks, thread scheduling, and varying work per transaction. Although there are a variety of tracing tools that provide some visibility into application internals (e.g., strace to gain visibility into I/O operations, and DTrace~\cite{brendan:dtrace} to profile performance), these tools do not directly report performance variation or identify outlying behavior.  
 Moreover, general-purpose tracing tools introduce substantial (sometimes order-of-magnitude) slowdowns, 
 when collecting fine-grain measurements. For example, we report the overhead of DTrace in Section~\ref{ssec:vprofiler}. 
The overhead of these  tools skews application behavior and obscures millisecond- and sub-millisecond-scale root causes of latency variance.
In this section, we introduce \vprofiler, a novel tool for automatically instrumenting an application's source code to sample and profile execution time variance
at fine time scales with minimal overhead, preserving the behavior of the system under study.

 
\subsection{Variance Tree}

We can gain insight into why latency variance arises in an application by subdividing and attributing execution time across a call graph, similar to a conventional execution time profile generated by tools such as \texttt{gprof}.
However, rather than identifying functions that represent a large fraction of execution time, we instead calculate the variance and covariance of each component of the call graph across many invocations to identify those functions that contribute to performance variability.  
Two key challenges arise in this approach: (1) managing the hierarchical nature of the call graph and the corresponding hierarchy that arises in the variance of execution times, and (2) ensuring that profiling overhead remains low.  
We first discuss the former challenge, and address the latter in Section~\ref{sec:vprofiler}.

A variance tree is rooted in a function that is invoked repeatedly over the course of an application.  We measure latency and its variance across invocations.  For example, an a conventional DBMS architecture, where transactions are mapped to worker threads for execution, we examine the variance tree rooted in the dispatch function invoked in the worker thread's main loop to begin executing a new transaction.  In the case of MySQL, this function is \texttt{dispatch\_command}, and for Postgres it is \texttt{PostgresMain}.

\begin{figure*}
\centering
\inv\sinv
\includegraphics[scale=0.4]{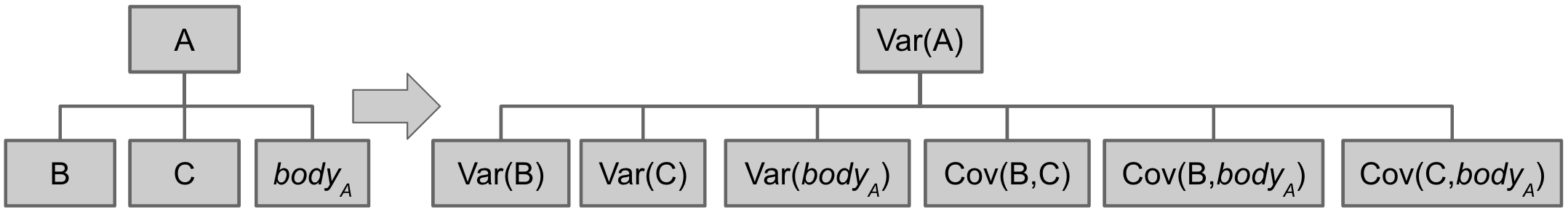}
\caption{A static call graph and its corresponding \textit{variance tree} (here, $body_A$ represents the time spent in the body of A).}
\inv\sinv
\label{figure:var_break_down}
\end{figure*}

Figure~\ref{figure:var_break_down} (left) depicts a sample call graph comprising a function A that invokes two children B and C and includes execution time in the body of A.
We can label each node in a particular invocation of this call graph with its execution time, yielding the relationship that the execution time of the parent node is the sum of its children, for example:
\begin{displaymath}
E(A) = E(B) + E(C) + E(body_A)
\end{displaymath}

We deconstruct and represent the variance of the call graph using the expression:
\begin{equation}
Var(\sum_{i=1}^{n}X_i) = \sum_{i=1}^{n}Var(X_i) + 2\sum_{1\leq i}\sum_{\leq j \leq n}Cov(X_i, X_j)
\label{eq:var-break-down}
\end{equation}
Figure~\ref{figure:var_break_down} (right) shows a corresponding visualization of the variances and covariances in a representation we call a \emph{variance tree}.

The variance tree allows us to quickly identify sub-trees that do not contribute to latency variability, as their variance is (relative to other nodes) small.
Identifying the root causes of large variance, however, is not so trivial.
The variance of a parent node is always larger than any of its children, so simply identifying the nodes with the highest variance is not useful for understanding the cause of that variance.
Furthermore, some variance arises because invocations may perform more work and manipulate more data (e.g., a transaction access more records). 
Such variance is not an indication of a mitigable pathology as the variance is inherent; our objective is to identify sources of variance that reveal performance anomalies and lead to actionable optimization opportunities.
High covariance across pairs of functions can be an indicator of a correlation between the amount of work performed by such functions.

At a high level, our goal is to use the variance tree to identify functions (or co-varying function pairs) that (1) account for a substantial fraction of overall latency variance and (2) are informative; that is, functions where analyzing the code will reveal insight as to why variance occurs.
To unify terminology, we refer to the variance of a function or co-variance of a function pair as a \emph{factor}.

Identifying factors that account for a large fraction of their parents' variance is straightforward.  
What is more complicated is how to identify functions that are informative.
We address this question in the next section.

\subsection{Ranking Factors}

Our intuition is that functions deeper in the call graph implement narrower and more specific functionality, and hence are more likely to reveal the root cause of latency variance.  
For example, consider a hypothetical function \texttt{WriteLog} that writes several log records to a global log buffer, but must first acquire the lock on the log buffer (\texttt{Lock}), copy the log data to the log buffer (\texttt{CopyData}), and finally release the lock (\texttt{Unlock}). Suppose \texttt{WriteLog}'s variance accounts for 30\% of its transaction latency variance, but \texttt{CopyData}'s accounts for 28\%. Analyzing \texttt{CopyData} is likely more informative even though it accounts for slightly less variance than \texttt{WriteLog}, because its functionality is more specific.  Further investigation may reveal the variance arises due to the size of log data being copied, suggesting mitigation techniques that reduce log size variance.

Based on this intuition, \vprofiler\ ranks factors using a score function that considers both the magnitude of variance attributed to the factor and its relative position in the call graph.
A particular factor may appear in a call graph more than once if a function is invoked from multiple call sites.
When ranking factors, \vprofiler\ aggregates the variance/covariance across all call sites.  

To quantify  a factor's  position within the call graph, \vprofiler\ assigns each function a height based on the maximum depth of the call tree beneath it.
For factors  representing the covariance of two functions, \vprofiler\ uses the maximum height of the two functions.
It uses a specificity metric that is a decreasing function of the height of a factor $\phi$: 
\begin{equation}
specificity(\phi) = (height(call\_graph) - height(\phi))^2
\label{eq:speci}
\end{equation}
where $height(call\_graph)$ is the height of the root of the call graph, and $height(\phi)$ is the height of the factor.

Similarly, \vprofiler\ uses a score function that jointly considers specificity and variance:
\begin{equation}
score(\phi) =  specificity(\phi)\sum_{i}V(\phi_i)
\label{eq:resp}
\end{equation}
where $V(\phi_i)$ represents the variance or covariance of a specific instance (call site) of a factor within the call graph.

\begin{algorithm}[t]
    \SetAlgoLined
    \SetKwInOut{Input}{Inputs}
    \SetKwInOut{Output}{Output}
    \SetKwFunction{FactorOf}{factor\_of}
    \SetKwFunction{FindFactor}{find\_factor}
    \SetKwFunction{speci}{specificity}
    
    \Input{
        $t$: variance break-down tree,\\
        $k$: maximum number of functions to select,\\
        $d$: threshold for minimum contribution}
        
    \Output{
        $s^*$: top $k$ most responsible factors}
    \BlankLine
        
    $h \gets $ empty list\;
    \BlankLine    
    
    \ForEach{$ node~ \phi \in t$} {
        $\phi^* \gets \FactorOf{$\phi$}$\;
        \uIf{$\phi^* \not\in h$} {
            $\phi^*.contri \gets t.contri$\;
            $h \gets h \cup \phi^*$\;
        }
        \uElse{
            $\phi'.contri \gets \phi'.contri + \phi.contri$\;
        }
    }
    \BlankLine    
    
    \ForEach{$\phi \in h$} {
        $\phi.score = \speci{$\phi$} \cdot \phi.contri$\;
    }
    \BlankLine
    
    Sort $h$ in descending order of $\phi.score$\;
    \BlankLine
    
    $s^* \gets$ empty list\;
    \For{$ i \gets 1$ \KwTo $k$} {
        $\phi \gets h[i]$\;
        \uIf{$\phi.contri \geq d$} {
            $s^* \gets s^* \cup \phi$\;
        }
    }
    
    \KwRet{$s^*$}\;
    
\caption{Factor Selection}
\label{alg:factor-selection}
\end{algorithm}

Given the \textit{variance tree}, we now describe an algorithm to select the top-k factors based on their score. 
The pseudocode is shown in Algorithm~\ref{alg:factor-selection}. 
For each node in the tree, we determine if the corresponding factor is already in list $h$. If not, we insert the factor and its (co-)variance into $h$. 
Otherwise, we accumulate the (co-)variance represented by the node into the existing element in $h$ (line 1 to line 10). 
Once we have calculated to total (co-)variance of each factor, we calculate their \textit{score} values using Equation~\ref{eq:resp} (line 11 to 13). Then, we sort factors in descending \textit{score} order, and select the top $k$  with a total (co-)variance greater than a threshold $d$ (line 14 to 23).

\subsection{Iterative Refinement}
\label{sec:vprofiler}

\begin{algorithm}[t]
    \SetAlgoLined
    \SetKwInOut{Input}{Inputs}
    \SetKwInOut{Output}{Output}
    \SetKwFunction{VarBreakDown}{var\_break\_down}
    \SetKwFunction{SetRoot}{set\_root}
    \SetKwFunction{AddChildren}{add\_children}
    \SetKwFunction{SelectFactors}{select\_factors}
    \SetKwFunction{NeedsBreakDown}{needs\_break\_down}
    \SetKwFunction{GetAnyInstanceOf}{get\_any\_instance\_of}
    \SetKwFunction{IsVariance}{is\_variance}
    \SetKwFunction{Clear}{clear}
    \SetKwFunction{Mark}{mark\_as\_selected}

    \Input{
        $v$: the starting function (i.e., entry point),\\
        $k$: maximum number of functions to select,\\
        $d$: threshold for minimum contribution}
        
    \Output{
        $s^*$: top $k$ most responsible factors}
    \BlankLine
        
    $t \gets$ tree with $Var(v)$ as root\;
    $l \gets \{Var(v)\}$\;
    $e \gets true$\;
    \BlankLine
    
    \While{$e$} {
        \BlankLine
        
        \ForEach{$factor~ f \in l$} {
            \If{$\IsVariance{f}$} {
                $c \gets \VarBreakDown{f}$\;
                $t.\AddChildren{f, c}$\;
            }
        }
        \BlankLine
        
        $s^* \gets \SelectFactors{t, k, d}$\;
        \BlankLine
        
        $l.\Clear{}$\;
        $e \gets false$\;
        \ForEach{$factor~ f \in s^*$} {
            \uIf{$\NeedsBreakDown{f}$} {
                $l \gets l \cup f$\;
                $e \gets true$\;
            }
            \uElseIf{$\IsVariance{f}$} {
                $\Mark{f}$\;
            }
        }
        \BlankLine
    }
    
    \KwRet{$s^*$}\;
    
\caption{Workflow of \vprofiler}
\label{alg:iterative-selection}
\end{algorithm}

Given a complete variance tree, factor selection (Algorithm~\ref{alg:factor-selection})  identifies
 the top factors that a developer should investigate further to identify the root causes of transaction latency variance.
However, collecting a complete variance tree is infeasible due to the enormous size and complexity of call graphs in modern DBMS software, such as MySQL and Postgres.
Instrumenting each function adds overhead to execution time, and if this overhead is too large, the variance tree is no longer representative of unprofiled execution.

Hence, \vprofiler\ iteratively refines  the profiling instrumentation to build a variance tree starting from the root of the variance tree until the profile is sufficient for a developer to identify key sources of variance.
In each iteration, \vprofiler\ identifies the top k factors when profiling a subset of functions, starting at the root of the call graph.
This profile is then returned to the developer, who determines if the profile is sufficient.  
If not, the children of the top-k factors are added to the list of functions to be profiled, instrumentation code is automatically
 inserted by \vprofiler, and a new profile is collected.
In detail:

\vspace{0.1cm}
\noindent\textbf{Initialization (Algorithm \ref{alg:iterative-selection}, line 1 to 3)}

\vprofiler\ starts with an empty variance tree, and initializes the list of functions to profile to contain only the root.

\vspace{0.1cm}
\noindent\textbf{Variance Break Down (Algorithm \ref{alg:iterative-selection}, line 5 to 8)}

For each profiled function, \vprofiler\ automatically instruments the code to measure the latency of all invocations of the function and the latency of each child.  
The variance and co-variances of these children are added to the variance tree, thereby expanding the tree by one level. 

\vspace{0.1cm}
\noindent\textbf{Factor Selection (Algorithm \ref{alg:iterative-selection}, line 9 to 17)}

After the variance tree is expanded, \vprofiler\ performs factor selection to choose the top k highest scoring factors within the tree, which are then reported to the developer.  If the profile is insufficient, the developer requests another iteration, which adds the children of these top k functions to the list to be profiled.

Note that, ultimately, \vprofiler's output is heuristic.  It identifies code that contributes to variance, but a developer must analyze this code to determine if the variance is inherent or is indicative of a performance pathology. 

 \vprofiler\ uses a parser to automatically inject instrumentation code as a prolog and epilog to each function that is selected for profiling using a source-to-source translation tool and then recompiling the binary.  Our approach is similar to conventional profilers, such as \texttt{gprof}, except that \vprofiler\ instruments only a subset of functions at a time.

\section{Latency Variance in MySQL}
\label{sec:mysql}

In this section, we use \vprofiler\ to analyze the source code of MySQL 5.6.23, and 
	characterize the main sources of variance therein.
  Here, we report our findings using the TPC-C benchmark. However, in Section~\ref{sec:experiments}, 
  we evaluate our techniques using 5 different benchmarks (including TPC-C) with various degrees of complexity and contention.

We use the OLTP-Bench~\cite{oltpbenchmark} software suite to 
run the TPC-C workload under two configurations.
First, we study a 
128-warehouse configuration with a 30 GB buffer pool on a system with 2 Intel(R) Xeon(R) CPU E5-2450 processors and 2.10GHz cores. 
Second, we study a reduced-scale 2-warehouse configuration with a 128M buffer pool on a machine with 2 Intel Xeon E5-1670v2 2.5GHz virtual CPUs. 
The reduced-scale configuration exaggerates buffer pool contention, revealing latency sources that may arise
	in workloads with a working set significantly larger than the available memory.
We refer to these configurations as 128-WH and 2-WH, respectively.
In both cases, we use a separate machine to issue client requests to the MySQL server.

Table \ref{tab:source:mysql} summarizes the key variance sources in MySQL identified by \vprofiler.
Whereas MySQL has one of the most complex code bases with over 1.5M lines of code and 30K functions, 
	\vprofiler\  narrows down our search by automatically identifying a handful of functions that 
		contribute the most to the overall transaction variance.  
	This clearly demonstrates the value of \vprofiler: 
	we only need to manually inspect  these few functions to understand whether their execution time variance 
	is inherent or is caused by a performance pathology that can be mitigated or avoided. 
	   Next, we explain the role of each of the functions found by \vprofiler.

\begin{table}[t]
\inv
\small
\centering
\begin{tabular}{|c|c|c|} \hline
Config&Function Name & Percentage of \\
           &              & Overall Variance \\ \hline
128-WH&\texttt{os\_event\_wait} [A] &37.5\%\\ \hline
128-WH&\texttt{os\_event\_wait} [B]&21.7\%\\ \hline
128-WH&\texttt{row\_ins\_clust\_index\_entry\_low}&9.3\%\\ \hline
2-WH&\texttt{buf\_pool\_mutex\_enter}&32.92\%\\ \hline
2-WH&\texttt{img\_btr\_cur\_search\_to\_nth\_level}&8.3\%\\ \hline
2-WH&\texttt{fil\_flush}&5\%\\
\hline\end{tabular}
\caption{Key sources of variance in MySQL.}
\inv\inv
\label{tab:source:mysql}
\end{table}

\subsection{os\_event\_wait} 
\label{sec:mysql:sched}
MySQL includes its own cross-platform API for locks and condition 
 variables; \texttt{os\_event\_wait} is one of the central functions in this abstraction layer. 
This function is similar to the platform-specific \texttt{pthread\_cond\_wait} function on Linux, which is used to wait on a conditional variable. 
MySQL uses \texttt{os\_event\_wait} extensively for synchronization.
The implementation of \texttt{os\_event\_wait} yields little insight into why transaction execution is blocked.
Instead, we examine the context for the two most significant call sites that invoke \texttt{os\_event\_wait} (referred to as A and B 
in Table~\ref{tab:source:mysql}).  Both call sites occur in the execution of  \texttt{lock\_wait\_suspend\_thread}, which is used to put a thread to sleep when its associated transaction tries to acquire a lock on some data record, but must wait due to a lock conflict. 
These two specific call sites correspond to locks acquired during select and update statements, respectively.

The implication of this result is that variability of wait time for contended locks is the largest source of variance in MySQL running TPC-C.  
This finding motivates our idea of variance-aware transaction scheduling in Section~\ref{sec:vats}, 
which seeks to 
minimize overall wait time variance by optimizing the order in which locks are granted to waiting threads.

\subsection{row\_ins\_clust\_index\_entry\_low} 

The function \texttt{row\_ins\_clust\_index\_entry\_low} inserts a new data record into a clustered index, which is a critical step in the execution of insert operations.
\vprofiler\ reports that none of the children of this function exhibit a significant amount of variance to be selected by the factor selection algorithm. 
Instead, \vprofiler\ reports the main variance to arise in the body of \texttt{row\_ins\_clust\_index\_entry\_low} due to varying code paths taken based on the state of the index prior to the insert operation.
The variance here is inherent to the index mutation, not a performance pathology.

\subsection{buf\_pool\_mutex\_enter}
\label{sec:mysql:buf}
The function \texttt{buf\_pool\_mutex\_enter} is called when other functions access the buffer pool.
Similar to \texttt{os\_event\_wait}, this function is called from various sites. 
However, the call most responsible for its variance occurs in \texttt{buf\_page\_make\_young}, which is used to move a page to the head of the list 
 managing buffer page replacements. 
InnoDB replaces buffer pool pages using a variant of the least recently used algorithm.
The buffer page replacement order is maintained in a list, called \texttt{LRU}.  
Upon certain types of accesses, a page must be located and moved to the head of the \texttt{LRU} list. 
Threads must acquire a lock before accessing the LRU list; that lock is acquired in \texttt{buf\_pool\_mutex\_enter}. 
The variance in this function reflects the wait time while other threads are reordering the LRU list using \texttt{buf\_page\_make\_young}.
We therefore propose an alternative strategy to manage buffer pool replacements in Section~\ref{sec:llu}.

\subsection{btr\_cur\_search\_to\_nth\_level}
The role of this function is to traverse an index tree, placing a tree cursor at a given level, and leaving a shared or exclusive lock on the cursor page. 
A performance-critical loop in this function traverses from level to level  in the index tree, and its runtime varies with the depth to which the tree must be traversed.  
The variance here is inherent to the index traversal, not a performance pathology.

\subsection{fil\_flush}
\label{sec:mysql:flush}
In operating systems that use disk buffering to improve I/O performance, 
MySQL uses \texttt{fil\_flush}  to flush redo logs generated by a transaction. When disk buffering is enabled, disk I/O latency variance is exposed in this function (rather than \texttt{write} system calls).
The variance here is inherent to the I/O, but might be mitigated by logging to faster I/O devices, e.g.,~\cite{nvram-recovery-pavlo,nvram-logging-ryan,pelleywenisch13}.


\section{Latency Variance in Postgres}
\label{sec:psql}

In this section, we use \vprofiler\ to  analyze the source code of Postgres 9.6---another extremely popular DBMS. 
For Postgres, we use a server with 2 Intel(R) Xeon(R) CPU E5-2450 processors and 2.10GHz cores, and 
use a separate client machine. 
	In this section, we use the  TPC-C benchmark with a 32-warehouse configuration and a 30 GB buffer pool.
Table \ref{tab:source:psql} shows the top three functions in Postgres source code identified by \vprofiler\ as the main sources of variance (the top source dominates, accounting for 76.8\%). 

\begin{table}[t]
\small
\sinv
\centering
\begin{tabular}{|c|c|c|} \hline
Function Name & Percentage of Overall Variance\\ \hline
\texttt{LWLockAcquireOrWait}&76.8\%\\ \hline
\texttt{ReleasePredicateLocks}&6\%\\ \hline
\texttt{ExecProcNode}&5\%\\ \hline
\hline\end{tabular}
\caption{Key sources of variance in Postgres.}
\inv
\label{tab:source:psql}
\end{table}

\subsection{LWLockAcquireOrWait}
\label{sec:psql:src:logging}
When a transaction modifies a page, 
one or more redo log records are generated to record  its modifications.
Postgres uses write-ahead logging to achieve atomicity and durability;
before a transaction commits, all its  redo logs   must be  flushed to disk. 
 Postgres uses a single global lock object, called \texttt{WALWriteLock}, to ensure that only one transaction is flushing redo logs at a time. 
 In particular, each transaction must call the \texttt{LWLockAcquireOrWait} function  to acquire \texttt{WALWriteLock} exclusively before it can
 write and flush its redo log records. 
 The latency variance in \texttt{LWLockAcquireOrWait} arises due to varying wait times to acquire this lock.
  A natural solution is therefore to reduce contention for this global lock, or to allow 
  for multiple transactions to flush simultaneously. The former may be attempted by accelerating I/O (e.g., tuning the I/O block
  size in Postgres, or by placing the logs on a NVRAM~\cite{nvram-recovery-pavlo,nvram-logging-ryan} or SSD~\cite{bs-shimin,bs-KennethRoss}), whereas the  latter can be attempted by a variety of distributed logging schemes (e.g.,~\cite{dist-logging-old,partitioned-logging}). Both of these strategies have proven effective in improving throughput and mean latencies~\cite{nvram-recovery-pavlo,nvram-logging-ryan}.
		However,  \vprofiler's findings regarding \texttt{LWLockAcquireOrWait}'s contribution to the overall latency variance, 
			call for also vetting these strategies in terms of improving the predictability of Postgres performance. 
		We implement and study some of these ideas for Postgres  in Sections~\ref{sec:psql:sol:dist} and~\ref{sec:expr:psql}.   

\subsection{ReleasePredicateLocks}
Postgres uses predicate locking to avoid the phantom problem where a read conflicts with later inserts or updates adding new rows to the selected range of the read. 
As a transaction accesses rows in the database, locks are acquired on them to prevent other transactions from inserting new rows into its selected range. When the transactions commits (or rolls back), all its predicate locks are released by calling \texttt{ReleasePredicateLocks}. A variety of lock conflicts can be discovered upon release,
\tempcut{(e.g., RW-conflicts, out-conflicts and in-conflicts to committed transactions)}
and the execution time varies with the number and type of conflicts.
However, \texttt{ReleasePredicateLocks} accounts for only 6\% of overall variance, hence, we do not pursue it further.

\tempcut{
\subsection{ExecProcNode}
After parsing, Postgres generates an execution plan for each query. This plan is a tree-like structure with multiple plan nodes, such as scans, joins or materialization operations. Depending on the type of each node, the \texttt{ExecProcNode} function invokes a series of other functions, such as \texttt{ExecInitSeqScan} or \texttt{ExecInitNestLoop}, to perform the required work. The variance of \texttt{ExecProcNode} therefore stems from differences in query plans. No single child of this function accounts for a significant fraction of variance.}



\section{Variance-Aware Transaction \\Scheduling}
\label{sec:vats}

According to \vprofiler's findings from Section~\ref{sec:mysql:sched}, locks wait times can account for a 
significant portion of the overall latency variance (e.g., over 
59.2\% in MySQL based on Table~\ref{tab:source:mysql}).
Motivated by this finding, in this section we aim to design a lock scheduling algorithm that can dramatically reduce latency variance.


\subsection{Problem Setting}

Traditional databases often rely on variants of 2-phase locking (2-PL) for concurrency control.
A transaction may request locks on different database objects (e.g., rows) at different points in its lifetime in the system.
Conceptually, each database object $b$ has its own queue $Q_b$. When a transaction $T$ requests a lock on $b$, the lock is immediately granted if (i) no other locks are currently held on $b$ by other transactions, or (ii) the current locks on $b$ are compatible 
with the requested lock type and there are no other transactions currently waiting in $Q_b$.\footnote{To prevent the reads from starving the writes indefinitely,
	 newly arrived read requests are usually not granted if there are already write requests waiting in the same queue.}
	  For example, read and write locks are incompatible in a serializable database. 

However, when a lock on $b$ cannot be granted immediately, 
	transaction $T$ is suspended and placed in $Q_b$ to wait until its lock can be granted.
In general, each transaction may wait in multiple queues during its life time,
and each queue might contain multiple transactions waiting in it. 
Let $Q_b=\{T_1,\cdots,T_n\}$ denote the transactions currently waiting to be granted a lock on $b$. 

Now, once all the currently held locks on $b$ are released, the \emph{lock scheduling} (a.k.a. \emph{transaction scheduling})
problem is the decision regarding 
which transaction(s) in $Q_b$ must be granted their lock request next. 
The transaction scheduler might choose one of the exclusive (e.g., write) requests, 
	or choose one of more of the inclusive ones.

The default transaction scheduling in many databases (including MySQL and Postgres among others) is the First-Come-First-Served (FCFS) algorithm. In this algorithm, whenever a lock on $b$ becomes available, 
	the transaction which has arrived in $Q_b$ the earliest is granted the lock, say $T_e$.
		Additionally, if $T_e$ is inclusive, all the other transactions in $Q_b$ whose requests 
		are compatible with that of $T_e$ are also granted a lock. 
	In other words, $T_e$ is selected based on the amount of time it has spent in the current queue (not in the system).
Fairness and simplicity have contributed to FCFS's popularity. 
However, FCFS does not even minimize mean latency, let alone latency variance.

\begin{figure*}[t]
    \centering
    \includegraphics[width=0.4\textwidth]{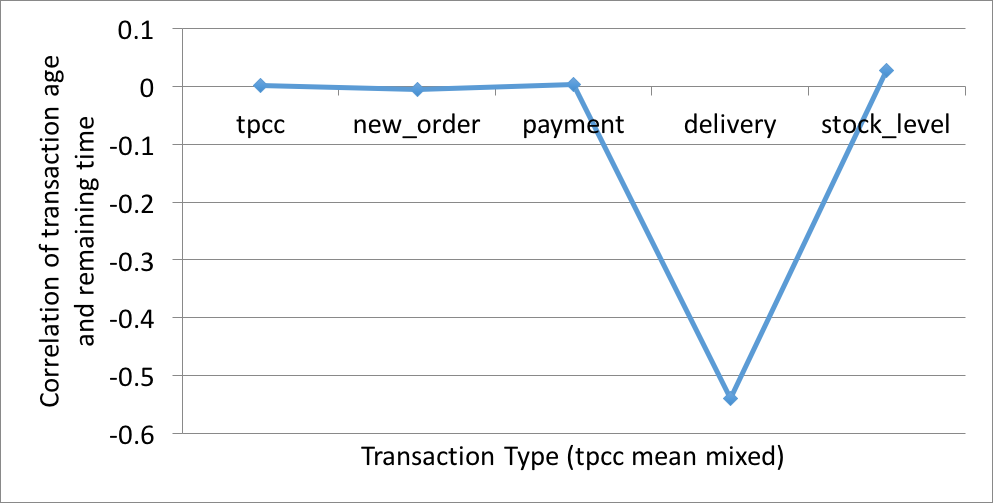}
    \caption{Correlation between transaction age and remaining time for different transaction types (TPC-C benchmark)}
    \label{fig:correlation}
\end{figure*}

\ph{Challenge of unpredictable remaining times}
One key challenge in devising effective transaction scheduling algorithms 
	is the lack of prior knowledge regarding a transaction's remaining time. 
	In other words, when a transaction arrives in $Q_b$, the system is only aware of its \emph{age} 
	(i.e., elapsed since its birth),
		but does not know long it will yet take to finish (and release its locks) once it is granted a lock on $b$.
		For example, it may need to wait in a few other locks before it can proceed to completion.
	In fact, figure \ref{fig:correlation} reveals that there is very little correlation between a transaction's age and its overall latency in practice (this holds even if one restricts  oneself to a particular class of  transactions, e.g., New Order or Delivery in TPC-C).
	Thus, when devising  a scheduling strategy, we must account for the fact that remaining times 
		are known and hard to estimate.

\subsection{A Convex Loss Function}
\label{sec:convex}

Our ultimate goal in this paper is to improve predictability by reducing latency variance and tail latencies.
 However, solely minimizing variance as a loss function may lead to undesirable side effects. For example, consider a 
 transaction scheduling 
algorithm that deliberately adds a large delay to every  completed transaction before allowing it to leave the system.
When a transaction's original latency is $l$, a sufficiently large delay $L\gg l$ will lead to variance of near zero, but will also
	significantly increase the mean latency, which is unacceptable in practice.
To exclude such scheduling strategy,
 	a more effective loss function is the so called $L_p$ norm, which if minimized, will indirectly reduce both deviations (e.g., variance or tail latencies)  and mean latency.  Formally,  when $n$ transactions finish with latencies
	 $\langle l_1,\cdots,l_n\rangle$, their $L_p$ norm (denoted as $||.||_p$) is defined as
\begin{equation}
	L_p = ||\langle l_1,\cdots,l_n \rangle ||_p= (\sum_{i=1}^n |l_i|^p)^{1/p}
\label{eq:lpnorm}
\end{equation}
where $p\geq 1$ is a real-valued number. Intuitively, the larger the $p$ value, the more we are penalizing  deviations of the $l_i$ values from mean. For example, as $p\rightarrow \infty$, $L_p$ norm approaches the max value of the list.
A typical value of $p$ in practice is 2.  However, our results in this section hold for all $p\geq 1$ values.

\subsection{Our VATS Algorithm}
\label{sec:grant}

Before presenting our algorithm, we need to define some notations.
Let $A(T)$ be the age of transaction $T$ when it arrives at a queue $Q_b$.
$Q_b$ is the set of transactions waiting to be granted a lock on $b$. 
  We define the history $H_b$ of an object $b$ to be the schedule of prior (and current) transactions holding a lock on $b$.
  In the following, we drop $b$ from our notation for convenience. 
  Let $F$ be some advice about the future, which an algorithm might have access to (our algorithm will not take such advice, but we will compare our algorithm to other algorithms that do have access to some advice).  

A scheduler $S = (S_f, S_a)$ is a set of two functions: $S_f, S_a: H \times Q \times F \rightarrow 2^{Q}$.
When the locks become available, the function $S_f$ determines which transactions from $Q$ should be granted a lock.
Note that $S_f$ cannot grant two exclusive locks on $b$ simultaneously.
When a new transaction arrives at $Q$, the function $S_a$ decides which transactions should be granted a lock.
Note that, when there are other locks currently held,  $S_a$ can only grant additional inclusive locks subject to their compatibility
 with the currently held locks.
 
Let $R(T)$ be a random variable indicating $T$'s remaining time once it is granted a lock on $b$.
Finally, let a menu $M$ be a sequence of transactions, where each transaction has an age and an arrival time when they arrive at the queue.  This will define a problem instance. 

We define the $p$-performance of a schedule $S$ on a menu $M$ to be the expected $L_p$ norm loss of the schedule $S$ on the menu $M$.   

\ph{Our Algorithm}
Given a menu, our goal is to design a scheduler that minimizes the expected $L_p$ norm loss.  To this end, we define our \vats\ scheduler
$S^{VATS} = (S_f^{VATS}, S_a^{VATS})$ as follows:
\begin{itemize}
  \item $S_f^{VATS}$ grants the lock to the eldest transaction, i.e., one with the largest age.
  \item $S_a^{VATS}$ never grants any locks.
\end{itemize}

In general, optimal scheduling is an $NP$-complete problem when the $R(T)$ values are known~\cite{pinedo2012scheduling}.
  Additionally, the on-line problem of scheduling even on one processor is impossible to do with a
   competitive ratio of  $O(1)$.\footnote{That is, for every scheduler $S$, there exists a menu $M$ where the optimal (off-line) algorithm performs $\omega(1)$ better than $S$.}

Interestingly, and counter-intuitively, in this paper we show that optimal scheduling becomes easier when the remaining times are not known!
Specifically, we avoid the above negative results by assuming that the $R(T)$s are I.I.D. random variables drawn from some (unknown)
distribution $D$.\footnote{To be more precise, what happens after transactions are granted a lock may depend on our schedule itself,
   as similar transactions could interact in the future on other queues. For simplicity, in this discussion we ignore this complication.}

 We now show that our \vats\ algorithm performs optimally,  even against algorithms that know the distribution $D$, 
  (i.e., algorithms that receive $F=D$ as an advice).
 Note that \vats\ does not use or need any distributional information or advice on future.  
 Interestingly, this will hold even if the menu and distribution are adversarially chosen.

\begin{theorem}
Fix any menu $M$, $p \geq 1$, and distribution $D$ with finite expected $L_p$ norm.  Let the $R(T)$s be i.i.d random variables drawn from  $D$.
Then the $p$-performance of \vats\ is optimal against all schedulers, even those that are given $D$ as advice about the future.
\label{thm:1}
\end{theorem}

\tempcut{
\cancut{Before we prove the theorem, we note that many stronger versions of the theorem are note true.
For example, our definition of a scheduler implicitly assumes that the processor is never idle when there are tasks in the queue.  This makes sense if the scheduler is ignorant about future transactions that will arrive in the following sense: the only reason to not schedule a task is because the scheduler anticipates that an important transaction will soon arrive and so is afraid of the opportunity cost of scheduling the task.  However, if the schedule does not know the future, the opportunity cost of scheduling will be equally great in the future.}


\begin{proof}
Assume for the sake of contradiction that there exists a menu $M$ of $\ell$ transactions $T_1, T_2, \ldots, T_{\ell}$ where a schedule $S$ has $p$-performance better than $S^{VATS}$.  We will transform $S$ into $S^{VATS}$ by a series of $\ell$ transformations: $S = S_0 \rightarrow S_1 \rightarrow S_2 \rightarrow \cdots \rightarrow S_{\ell}= S^{VATS} $.  We will show after each transformation that the performance of the schedule improves.  This yields a contradiction to the assumption that the $p$-performance of $S$ was better than that of  $S^{VATS}$.

In the $k$th transformation, we modify $S_{k-1}$ so that, if ever $S_{k-1}$ schedules a transaction $T_{k'}$ when $T_{k}$ is the eldest transaction in the queue, then $S_{k}$ will transpose the order of $T_k$ and $T_k'$, but otherwise run identically to $S_{k-1}$.

Note that $S_{\ell} = S^{VATS}$, because $S_{\ell}$ will run the eldest transaction, no matter which one it is.

Let $T_{S_{k-1}, 1}, T_{S_{k-1},2}, \ldots, T_{S_{k-1}, \ell}$ and $T_{S_{k}, 1}, T_{S_{k},2}, \ldots, T_{S_{k}, \ell}$  be the order of transactions scheduled in $S_{k-1}$ and $S_k$ respectively.   Note that these may be random variables, in that the $i$th transaction scheduled might depend on the randomness of the scheduler, as well as the time that previous transactions held onto the lock.  Let $U_{S}(T)$ be the time it takes between when $T$ arrives and when the lock is first free under schedule $S$.  Let $W_S(T)$ be the set of transactions scheduled while $T$ is in the queue (including $T$) under schedule $S$.

To compare the performance of $S_{k-1}$ and $S_k$, we create a coupling between two different drawings $D_1$ and $D_2$ of the $R(\cdot)s$ so that for all $i$, $R_{D_1}(T_{S_{k-1}, i}) = R_{D_2}(T_{S_{k}, i})$.  First note there is no dependency problem here because (by induction on $i$) under this coupling $T_{S_{k-1}, i}$ and $T_{S_{k}, i}$  will be schedule at the same time.  Also, because the $R(\cdot)$s are all drawn i.i.d, this is a valid coupling, which is to say that $D_1$ and $D_2$ are (marginally) drawn from the same distribution.

Note that the performance of $S_{k-1}$ and $S_k$ are respectively,
\begin{align*}
\int_{D_1} \left( \sum_i  |A[T_{S_{k-1}, i}]  +  U_{S_{k-1}}(T_{S_{k-1}, i}) \right. \\ \left. + \sum_{T_j \in W_{S_{k-1}}(T_{S_{k-1}, i})} R(T_j)|^p  \right)^{1/p}
\end{align*}
and
\begin{align*}
\int_{D_2} \left( \sum_i  |A[T_{S_{k}, i}] + U_{S_{k}}(T_{S_{k}, i}) + \sum_{T_j \in W_{S_{k}}(T_{S_{k}, i})} R(T_j)|^p  \right)^{1/p}
\end{align*}

To show that the first is greater than the second, we fix some realization of $D_1$.  Using our coupling, this gives us a realization of $D_2$.  We will show that no matter what the realization is we have:
\begin{align*}
&\sum_i  |A[T_{S_{k-1}, i}] + U_{S_{k-1}}(T_{S_{k-1}, i}) + \sum_{T_j \in W_{S_{k-1}}(T_{S_{k-1}, i})} R(T_j)|^p  \label{eq:1} \\ &<  \sum_i  |A[T_{S_{k}, i}] + U_{S_{k}}(T_{S_{k}, i}) + \sum_{T_j \in W_{S_{k}}(T_{S_{k}, i})} R(T_j)|^p 
\end{align*}

Note that the  summands are identical except, possibly, for the terms of $T_k$ and $T_{k'}$.  Let $W_k = W_{S_{k-1}}(T_k) \cap W_{S_{k}}(T_k)$ be the transactions scheduled before $T_k$ in both schedules.  Define $W_{k'}$ analogously.  Let $W'$ be the transactions scheduled between $k$ and $k'$.
Then,  $W_{S_{k-1}}(T_k) = W_k \cup \{T_{k'}\} \cup W'$,  $W_{S_{k-1}}(T_{k'}) = W_{k'}$,
$W_{S_{k}}(T_k) = W_k$, and  $W_{S_{k}}(T_{k'}) = W_{k'} \cup \{T_{k'}\} \cup W'$.

The rearrangement inequality states that if $x_1, x_2, y$ are all nonnegative numbers then $|x_1 + y|^p + |x_2|^p \leq |x_1|^p + |x_2 + y|^p$ if and only if $x_1 \leq x_2$.
By applying the rearrangement inequality where
$$x_1 = A(T_{k'}) + U_{S_{k-1}}(T_{k'}) + \sum_{T_j \in W_{k'}} R_{D_1}(T_j)$$
$$x_2 = A(T_k) + U_{S_{k-1}}(T_k) +  \sum_{T_j \in W_k} R_{D_2}(T_j)$$
and $$y = R_{D_1}(T_{k'}) + \sum_{T_j \in W'} R(T_j) = R_{D_2}(T_{k}) + \sum_{T_j \in W'} R(T_j).$$

Since $T_k$ is older than $T_{k'}$ and since $R_{D_1}(T_{k'}) = R_{D_2}(T_k)$,
we have that $x_1 < x_2$.

The theorem follows by noting that in the $S_{k-1}$ schedule, the $T_{k'}$ term is $x_1$ and the $T_{k}$ term is $x_2 + y$; while in the $S_k$ schedule, the $T_{k'}$ term is $x_1 + y$ and the $T_{k}$ term is $x_2$.
\end{proof}
}

\tempcut{It is an interesting problem to relax the requirements of our theorem to be $O(1)$-competitive, rather than optimal, and show that our theorem can be (or cannot be) extended to these settings.  However, this is beyond the scope of the present work.}

%

\subsection{Practical Considerations} 
\label{sec:practical}

In our implementation, we make a few modifications to the \vats\ algorithm described in Section~\ref{sec:grant}.
First, whenever a read lock is granted by the algorithm, we also grant other compatible locks in the queue.
Our intuition is that, if our algorithm is effective at reducing variance at an equilibrium, 
	we do not expect a few transaction to have significantly higher completion times
		than a single one.
  
Moreover, \vats\ can  incur an overhead to (i) sort lock requests by age upon each release operation, or (ii) maintain
	 a min-heap  upon each transaction arrival. 
	This overhead is not justified when lock contention is rare.
	Hence, we only activate our \vats\ scheduling when the  the fraction of wait locks is higher than a pre-selected threshold $R$,  otherwise FCFS is used. 
	We evaluate \vats\, and the impact of $R$ in Sections~\ref{ssec:alg} and~\ref{ssec:para-tuning}, respectively.

\section{Achieving Predictability via\\ DBMS-Specific Optimizations}
\label{sec:predict:specific}

Following our findings from Sections~\ref{sec:mysql} and~\ref{sec:psql}, 
	we present further strategies for improving performance predictability. 
Unlike our \vats\ algorithm which is DBMS-independent, the techniques in this section are DBMS-specific: MySQL (Section~\ref{sec:llu}), Postgres (Section~\ref{sec:psql:sol:dist}), or both (Section~\ref{sec:va-tuning}).

\subsection{Lazy LRU Update (LLU)}
\label{sec:llu}

As noted in Section~\ref{sec:mysql}, the lock on the LRU list
 is a main source of variance in MySQL when the working set to buffer pool ratio is high, e.g., in our 2-WH configuration.

Algorithm~\ref{alg:buf-page-make-young} shows the sequence of events in MySQL for updating the LRU list.
First, a  mutex is acquired by calling \texttt{buf\_pool\_mutex\_enter}, and then a buffer page is moved to the
head of the LRU list by calling \texttt{buf\_page\_make\_young}. 

To improve its cache performance, InnoDB does not implement the strict LRU policy.
Instead,  it splits the LRU list into two sublists, \emph{young} and \emph{old}. 
By default, 3/8 of the pages at the tail of the list are placed on the old list;
replacement victims are selected from this list.
Upon a page access, if the page is currently in the old list, it is moved to
the head of the young list, and the tail of the young list is placed at the head
of the old list.  
InnoDB does not maintain precise LRU ordering within the young list.
This optimization avoids frequent re-ordering of the LRU list when the database working set fits within 5/8 of the buffer pool, avoiding the need to frequently acquire the buffer pool lock.
However, if old pages are accessed frequently, the lock becomes a bottleneck.
Our idea is to further relax the precision of LRU tracking to avoid this contention, as described next.

 Our proposed algorithm, Lazy LRU Update (LLU), limits the time that \texttt{buf\_pool\_mutex\_enter}  waits for the lock to avoid excessive delays. In other words,  we replace the mutex to a spin lock in order to control the wait time.
Since this lock is typically uncontended when buffer pool capacity is sufficient, using a spin lock instead of a mutex introduces minimal overhead.
However, if a waiting thread is unable to acquire the lock within 0.01ms, we abandon the attempt to update the global LRU list.
We instead add the page to a thread-local backlog of deferred LRU updates, $l$.
Later, when  \texttt{buf\_pool\_mutex\_enter} successfully acquires the lock when attempting to manipulate the LRU list for a different page, we first process all pages in $l$ before moving the page that triggered the reordering.  
Note that we must confirm that each of these pages is still present in the buffer pool before adding it to the young list, as the page may since have been evicted. 

\begin{algorithm}[t]
\inv\sinv
    \SetAlgoLined
    \SetKwInOut{Input}{Inputs}
    \SetKwInOut{Output}{Output}
    \SetKwFunction{BufPoolMutexEnter}{buf\_pool\_mutex\_enter}
    \SetKwFunction{BufLRUMakeBlockYoung}{buf\_LRU\_make\_block\_young}
    \SetKwFunction{BufPoolMutexExit}{buf\_pool\_mutex\_exit}
    
    \Input{
        $p$: buffer page to be moved to the start of the \texttt{LRU} list,\\
        $b$: the buffer pool}
        
    \BlankLine
        
    $\BufPoolMutexEnter{b}$\;
    $\BufLRUMakeBlockYoung{b, p}$\;
    $\BufPoolMutexExit{b}$\;
    
\caption{Baseline LRU Update}
\label{alg:buf-page-make-young}
\end{algorithm}

\tempcut{
\begin{algorithm}[t]
    \SetAlgoLined
    \SetKwInOut{Input}{Inputs}
    \SetKwInOut{Output}{Output}
    \SetKwFunction{SpinForTime}{spin\_for\_time}
    \SetKwFunction{Append}{append}
    \SetKwFunction{Remove}{remove}
    \SetKwFunction{InBufferPool}{in\_buffer\_pool}
    \SetKwFunction{BufLRUMakeBlockYoung}{buf\_LRU\_make\_block\_young}
    \SetKwFunction{BufPoolMutexExit}{buf\_pool\_mutex\_exit}
    
    \Input{
        $p$: buffer page to be moved to the start of the \texttt{LRU} list,\\
        $b$: the buffer pool,\\
        $l$: list of pages failed to be moved,\\
        $t$: timeout for spin lock}
        
    \BlankLine
        
    $s \gets \SpinForTime{b, t}$\;
    \BlankLine
    
    \uIf{$s = failure$} {
        \If{$p \in l$} {
            $l.\Remove{p}$\;
        }
        $l.\Append{p}$\;
        Return\;
    }
    \uElse {
        \ForEach{$page~u \in l$} {
            \If{$\InBufferPool{b, u}$} {
                $\BufLRUMakeBlockYoung{b, u}$\;
                $l.\Remove{u}$\;
            }
        }
        $\BufLRUMakeBlockYoung{b, p}$\;
    }
    \BlankLine
    
    $\BufPoolMutexExit{b}$\;
    
\caption{Lazy LRU Update}
\label{alg:buf-page-make-young-spin}
\end{algorithm}
}

\subsection{Distributed Logging}
\label{sec:psql:sol:dist}

As discovered by \vprofiler\ in Section~\ref{sec:psql},  over 70\% of latency variance in Postgres is due to
 the variation of wait times in redo log flush operations. 
 Thus,  a natural approach  to improving predictability is to use distributed logging, so that when a set of log files is unavailable, a transaction can write to another set of files instead of having to wait. 
 There are sophisticated schemes for distributed logging~\cite{dist-logging-old,partitioned-logging}.
Here, we implement a simple variant that allows Postgres to use two hard disks for storing two sets of redo logs.
A transaction only needs to wait when neither of these two sets is available, in which case it waits for the one with fewer waiters. While distributed logging is well-studied for improving mean latencies, our goal is to vet its effectiveness 
in reducing latency variance in Section~\ref{sec:expr:psql}.

\subsection{Variance-Aware Tuning}
\label{sec:va-tuning}

Our profiling results in Sections~\ref{sec:mysql} and~\ref{sec:psql} identified functions that account for substantial variance,
 whose behavior depend heavily on tuning parameters in MySQL or Postgres. 
 In this section, we discuss these parameters (we empirically examine their impact on latency variance in Section~\ref{sec:experiments}). 

First, from our investigation of \texttt{buf\_pool\_mutex\_enter} (Section~\ref{sec:mysql:buf}), we learned that buffer pool capacity (relative to the database working set) substantially impacts variance (and obviously, mean latency).
Hence, we sweep buffer pool capacity from 33\% to 100\% of the overall database size and measure impact on transaction variance.

Second, we learned that MySQL's policy regarding log flushing 
 has a noticeable influence on transaction variance (Section~\ref{sec:mysql:flush}).
MySQL's use of buffered I/O for redo logs involves two steps: a \texttt{write} system call, and a \texttt{flush} system call.
MySQL offers three policies through the
the \texttt{innodb\_flush\_log\_at\_trx\_commit} parameter:

\mph{Eager flush} This requires that redo logs are written and flushed by the transaction worker thread before committing the transactions. 
 
\mph{Lazy flush}  Under this setting, redo logs are written by the transaction worker thread, but flush operations are deferred to a separate log flusher thread, which invokes the flush system call roughly once per second.  Transactions may commit before their logs are flushed.

\mph{Lazy Write}  Under this setting, redo logs are prepared but not written by the transaction worker thread.  Both writing and flushing the log are deferred to a log flusher thread, which, again performs these operations once per second.  Transactions may commit before their logs are written.

Note that both lazy flush and lazy write risk losing forward progress in the event of a crash; transactions executed in the previous second may be reported as committed to the user, but may be unrecoverable because their redo logs never became durable.
Nevertheless, in contexts where forward progress loss can be tolerated, employing lazy flushing and writes can substantially improve the latency and predictability of transaction execution.

Finally, we observed that much of the latency variance in Postgres is due to varying wait times of 
	transactions when flushing their redo logs (Section~\ref{sec:psql:src:logging}). 
	This I/O operation can be accelerated by tuning Postgres's block size parameter, which is by default 8 KB. 
	(Another solution is to use distributed logging; see Section~\ref{sec:expr:psql}).


\section{Experiments}
\label{sec:experiments}

Our experiments aim to answer the following  questions:
\begin{enumerate}[leftmargin=0.2cm]
\item How effective is our \vats\ algorithm in reducing tail latency and latency
	variance compared to other lock scheduling algorithms? 
	How effective are our variance-aware tuning and other DBMS-specific strategies
	 in this regard?  What is the combined impact of all these strategies on reducing latency variance?

\item  Does our reduction of latency variance come at the cost of sacrificing mean latency or throughput? 

\item How effective and efficient is \vprofiler\ compared to other profiling tools and alternatives? 
\end{enumerate}
In summary, our experiments indicate the following:

\begin{itemize}[leftmargin=0.2cm]
\item For contended workloads (TPC-C, SEATS, and TATP), 
our \vats\ algorithm significantly improves upon FCFS (the scheduling used by MySQL, Postgres, and others),
reducing mean, variance, and 99th percentile latencies on average by 
	26\%, 37\%, and 36.8\%, respectively (and up to 59.7\%, 64\%, and 64.4\%, resp.)
	 without compromising throughput. 
	As expected, for non-contended workloads (Epinions and YCSB), the choice of scheduling algorithm is  
		immaterial. (Section~\ref{ssec:alg})

\item Our Lazy LRU Update (LLU) algorithm reduces MySQL's 
mean latency by 12.1\%, variance by 35.3\%, and 99th percentile latency by 26.2\%. (Section~\ref{ssec:spinlock})

\item Variance-aware tuning can also dramatically reduce the variance of latencies, depending on  memory  availability and durability requirements. (Section~\ref{ssec:tuning})

\item Given \vprofiler's findings on sources of variance in Postgres, we  
explore 
distributed logging, which reduces mean latency, variance and 99th percentile by 58.8\%, 44\% and 23.6\%, respectively. 
Likewise, choosing an appropriate  block size for redo logs reduces variance by 12.8\%. (Section \ref{sec:expr:psql})

\item \vprofiler's profiling overhead is an order of magnitude lower than  DTrace, and its factor selection algorithm reduces the number of required runs by several 
orders of magnitude compared to a na\"ive strategy.
(Section~\ref{ssec:vprofiler})

\item Our \vats\ algorithm can adaptively choose its own parameter value by observing the current variance of transaction latencies in the system. (Section~\ref{ssec:para-tuning})

\end{itemize}

Before presenting our results, we first introduce our experimental setup in Section~\ref{ssec:setup}.

\subsection{Experimental Setup} 
\label{ssec:setup}
The hardware and software used for our experiments in this section are identical to those described in Sections~\ref{sec:mysql} and~\ref{sec:psql}.
For a fair experiment, we used the same throughput of 500 transactions per second, across all workloads and algorithms.
Moreover, to rule out the effect of external load changes on latency variance, we used the OLTP-Bench~\cite{oltpbenchmark} tool to sustain a constant throughput
	throughout the experiment, and measured mean latency, latency variance, and 99th percentile latency for each algorithm and workload.
In addition to TPC-C, we also used the following workloads for a more extensive evaluation:

\mph{SEATS \cite{seats}} This benchmark is a simulation of an airline ticketing system using which customers search for flights and make online reservations. In our experiments, we used a scale factor of 50, leading to  a highly contended workload.

\mph{TATP \cite{tatp}} TATP is an OLTP application modeling a typical caller location system, used by tele-communication providers. 
For TATP, used a scale factor of 10 in our experiments. For this configuration, TATP is a contended workload (but not as contended as TPC-C).

\mph{Epinions \cite{epinions}} Epinions simulates a customer review website where users interact with each other and write reviews for various products. We used a scale factor of 500 in our experiments. This workload also has a very low contention.

\mph{YCSB \cite{ycsb}} YCSB is a set of micro-benchmarks simulating data management applications that have simple workloads but require high scalability. The scale factor used was 1200,  causing  little or no contention.

Given that varying lock wait times is a major problem for MySQL, we evaluate \vats\ in 
 Sections~\ref{ssec:alg} and~\ref{ssec:para-tuning} using MySQL. We also use MySQL in Sections~\ref{ssec:spinlock} and~\ref{ssec:tuning} since LLU and most of our variance-aware tuning  strategies apply to MySQL. Likewise, we use Postgres in 
 Section~\ref{sec:expr:psql} to evaluate variance reduction strategies for redo logs.

When the results are similar across all workloads, we only report the numbers for TPC-C as a representative workload. 

\ignore{
\begin{figure*}
    \centering
    \includegraphics[width=\textwidth]{plots/all}
    \caption{\tofix{Effect of combining all methods for reducing performance variance.}}
    \label{fig:all}
\end{figure*}

\subsection{\tofix{Combining All Methods} \label{ssec:combi}}
In this section, we show the best result we can have by combining each and every method we have for reducing the variance of transaction latency. We implement the algorithm described in Section~\ref{ssec:vats} in MySQL and set the value of \texttt{innodb\_flush\_log\_at\_trx\_commit} to 2 and the value of \texttt{innodb\_buffer\_pool\_size} to 40GiB, which is the maximum buffer pool size we can have. On the other hand, the original experiment has a value of 1 for \texttt{innodb\_flush\_log\_at\_trx\_commit} and a 7GB buffer pool. Figure \ref{fig:all} shows the improvement in performance and variance we get by combing all these methods. The combined method outperforms the original MySQL both in performance and variance of transaction latency. To be more precise, the combined method produces a 67.4\% reduction in mean latency and a 65\% reduction in variance.
}

\begin{figure*}[t]
    \centering
    \includegraphics[width=0.9\textwidth]{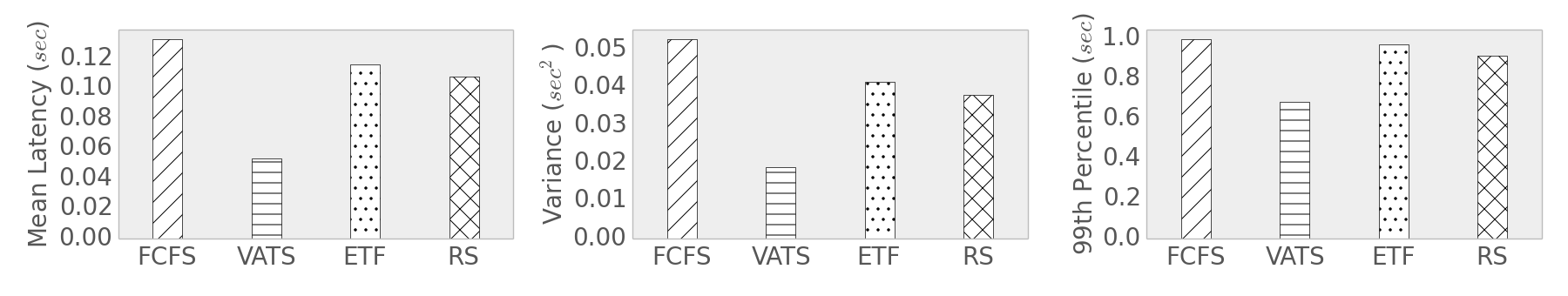}
\inv
    \caption{Effect of different scheduling algorithms on MySQL's performance (TPC-C benchmark).}
\inv
    \label{fig:sche-tpcc}
\end{figure*}

\begin{figure*}[t]
    \centering
    \includegraphics[width=0.9\textwidth]{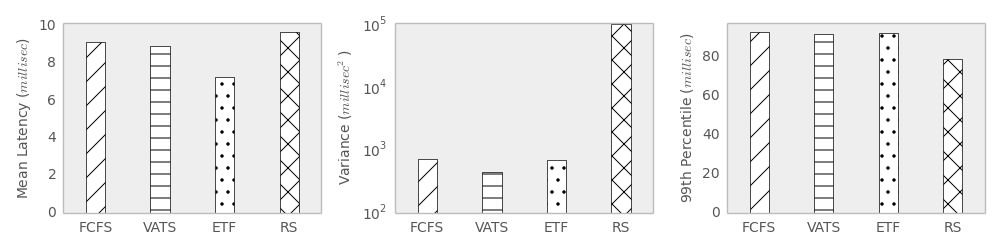}
\inv
    \caption{Effect of different scheduling algorithms on MySQL's performance (SEATS benchmark).}
\inv
    \label{fig:sche-seats}
\end{figure*}

\begin{figure*}[t]
    \centering
    \includegraphics[width=0.9\textwidth]{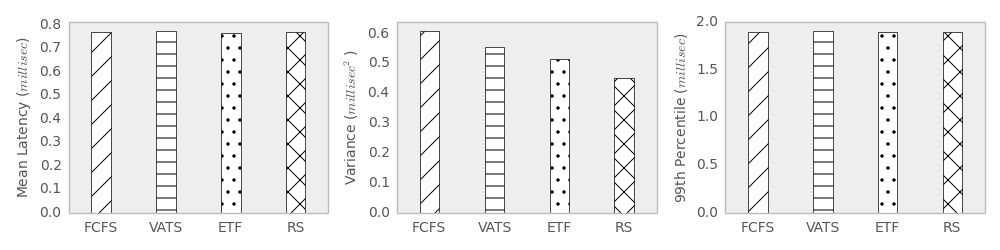}
\inv
    \caption{Effect of different scheduling algorithms on MySQL's performance (TATP benchmark).}
\inv
    \label{fig:sche-tatp}
\end{figure*}

\begin{figure*}[t]
    \centering
    \includegraphics[width=0.9\textwidth]{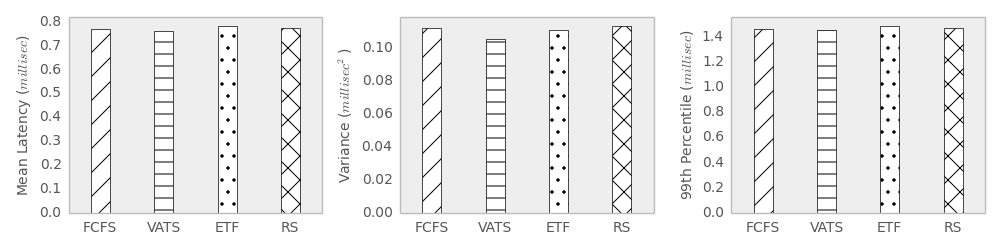}
\inv
    \caption{Effect of different scheduling algorithms on MySQL's performance (Epinions benchmark).}
\inv
    \label{fig:sche-epinions}
\end{figure*}

\begin{figure*}[t]
    \centering
    \includegraphics[width=0.9\textwidth]{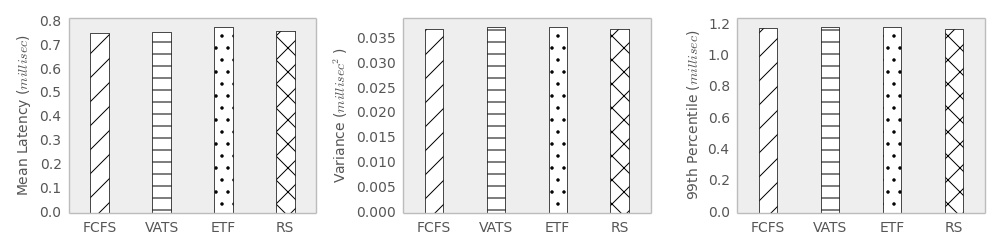}
\inv
    \caption{Effect of different scheduling algorithms on MySQL's performance (YCSB benchmark).}
\inv
    \label{fig:sche-ycsb}
\end{figure*}

\subsection{Studying Different Scheduling Algorithms} 
\label{ssec:alg}
In this section, we compare VATS to three other scheduling algorithms: 

\mph{First Come First Served (FCFS)}
This is the default scheduling in many DBMSs (including MySQL \& Postgres). 

\mph{Eldest Transaction First (ETF)}  Current transactions are sorted in the decreasing order of their age (i.e., the time 
since their were started), 
and their requested locks  are granted in this order, skipping those that are incompatible with the currently granted locks. 
This is similar to \vats, but without the adaptive strategy explained in Section~\ref{sec:practical}.

\mph{Randomized Scheduling (RS)}
Similar to ETF, except that instead of sorting transactions according to their age, they are sorted according to a random order.

The comparison is shown in Figure~\ref{fig:sche-tpcc}, \ref{fig:sche-seats}, \ref{fig:sche-tatp}, \ref{fig:sche-epinions} and \ref{fig:sche-ycsb}. In summary, the results indicate
that FCFS is a bad option for the three contended workloads. 
For example, for TATP, even a random scheduling (RS) improves upon FCFS by 25\% in terms of latency variance.
 However, the randomness of RS could also be a harm. For SEATS, RS is performing about 2 orders of magnitude worse than
 the rest of the scheduling algorithms.
Also, as expected, the choice of lock scheduling algorithm does not make a difference 
for Epinions and YCSB, simply because these workloads do not have any lock contention in the first place.


We have summarized \vats's improvement over FCFS in Table~\ref{tab:vats:summary} for all  workloads.
Our \vats\ algorithm is consistently superior for contended workloads and comparable to no-contention ones. 
On average, \vats\ reduces variance by 37.3\% across  all contended workloads, and 23\% across all five workloads. 
Most notably, \vats\ reduces the variance of TPC-C transaction latencies by 64\%.

\begin{table}[t]
\sinv\inv
\small
\centering
\newcolumntype{M}{>{\centering\arraybackslash}m}
\begin{tabular}{|b{0.1cm}|M{1.1cm}|M{1.1cm}|M{1.1cm}|M{1.1cm}|M{1.1cm}|} \hline
 & Workload & Mean    & 99th        & Variance & L2  \\
 &          & Latency & Percentile  &          & Norm \\ \hline
\multirow{4}{*}{\rotatebox{90}{Contended}} & TPCC & 59.7\% & 64.4\% & 64\% & 70\% \\ [.5ex] \cline{2-6}
                  & SEATS & 24.4\% &37.4\% & 37\% & 34\% \\ [.5ex] \cline{2-6}
                  & TATP & -6\% & 8.5\% & 11\% & 4\% \\ [.5ex] \cline{2-6}
                  & \textbf{Avg} & \textbf{26\%} & \textbf{36.8\%} & \textbf{37\%} & \textbf{36\%} \\ [.5ex] \hline
\multirow{4}{*}{\rotatebox{90}{No Contention\hspace{-\normalbaselineskip}}} & Epinions & 2.8\% & 1\% & 4\% & 2\% \parbox{0pt}{\rule{0pt}{6ex+\baselineskip}} \\ \cline{2-6}
                  & YCSB & -5\% & -1\% & -1\% & 0\% \parbox{0pt}{\rule{0pt}{6ex+\baselineskip}} \\
\hline\end{tabular}
\caption{\vats's relative reduction of mean latency, variance, 99th percetile and $L_2$ norm, compared to MySQL's FCFS lock scheduling, across different workloads.}
\inv\sinv
\label{tab:vats:summary}
\end{table}

\subsection{Lazy LRU Update Algorithm}
\label{ssec:spinlock}
\begin{figure*}[t]
\inv\sinv
    \centering
    \begin{subfigure}[t]{0.3\textwidth}
        \includegraphics[scale=0.3]{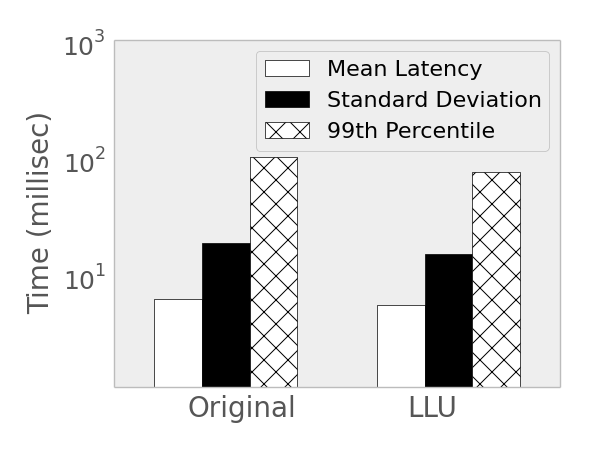}
    \end{subfigure}
    ~
    \centering
    \begin{subfigure}[t]{0.3\textwidth}
        \includegraphics[scale=0.3]{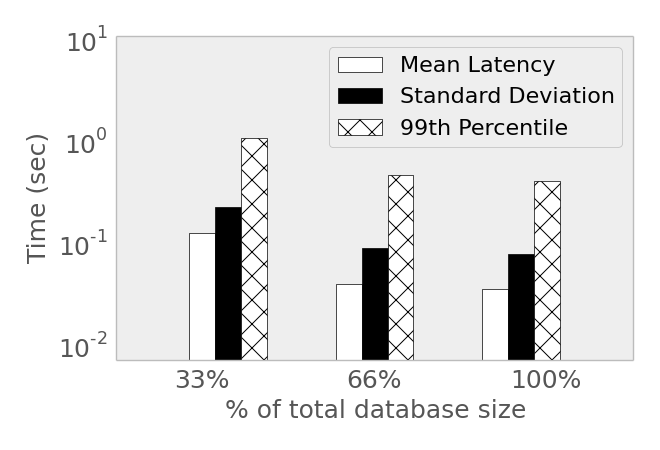}
    \end{subfigure}
    ~
    \centering
    \begin{subfigure}[t]{0.3\textwidth}
        \includegraphics[scale=0.3]{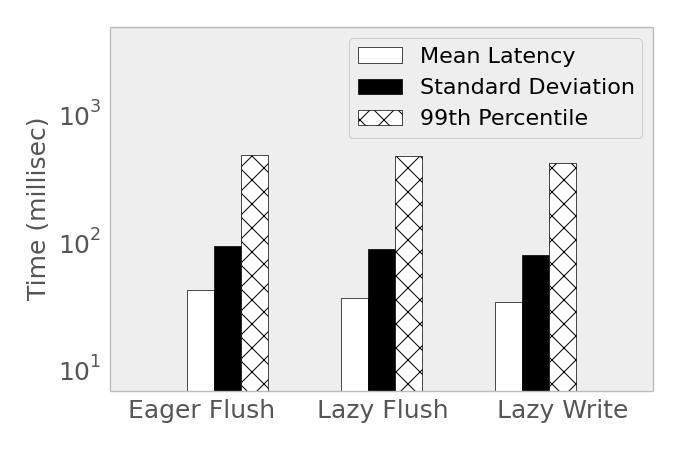}
    \end{subfigure}
\sinv
    \caption{Effect of LLU, buffer pool size (in \% of the entire database size), and log flush policy on MySQL (TPC-C).}
    \inv
    \label{fig:mysql_all}
\end{figure*}

In this section, we evaluate our Lazy LRU Update (LLU) algorithm, introduced in Section~\ref{sec:llu}.
To produce a memory-contended workload, we use the same 2-WH configuration of TPC-C as in Section~\ref{sec:mysql}.
As is shown in Figure~\ref{fig:mysql_all}(left), LLU improves  
mean latency by 12.1\%, variance by 35.3\%, and 99th percentile latency by 26.2\%.
This considerable improvement is because our LLU algorithm works by avoiding extremely long waits and delaying the operation of moving the buffer pages until its overhead is fairly cheap. This reduces the contention on the LRU data structure for memory-contended workloads.

\subsection{Variance-Aware Tuning} 
\label{ssec:tuning}
In Section~\ref{sec:va-tuning}, we identified several configuration parameters of MySQL, which our case study identified as having a large impact on transaction latency variance.

We first investigate the side of the buffer pool. The experiment results for TPC-C are shown in Figure~\ref{fig:mysql_all}(center).
 \ignore{(The results for other workloads are similar  and deferred to Appendix~\ref{app:expr:tuning}.)}
  We set the size of the buffer pool to 33\%, 66\%, and 100\% of the overall database size, which is 15GB. 
  As expected, increasing the buffer pool size will retain more data in memory, thus effectively reducing the number of buffer page evictions, the number of I/O operations, and 
   the degree of contention within the buffer pool. 
   As shown in Figure~\ref{fig:mysql_all}(center), 
   	the larger the size of the buffer pool, the lower the mean latency, the variance, and the 99th percentile latency. 
	Ideally, 
	choosing a buffer pool as large as the entire database size is recommended both for better average performance and for more predictability.
	However, depending on the working set size, smaller buffer pools might produce comparable results, e.g., in our experiments 66\% of the 
	database size seems more of an economical alternative. 

Second, we investigate MySQL's log flushing policies.
Figure~\ref{fig:mysql_all}(right) shows the experimental results for TPC-C. \ignore{The results for other workloads are similar (see Appendix~\ref{app:expr:tuning}).}
The results indicate that deferring both write and flush operations to a log flusher thread minimizes transaction variances.  
This result is not surprising: eagerly flushing logs prior to commit places highly variable disk write latencies on the transaction execution critical path.
As previously noted, however, lazy flushing introduces the risk that forward progress (committed transactions) may be lost in the event of a crash.

\subsection{Improving Predictability in Postgres}
\label{sec:expr:psql}

\begin{figure}[t]
    \centering
    \begin{subfigure}[t]{0.2\textwidth}
        \includegraphics[scale=0.35]{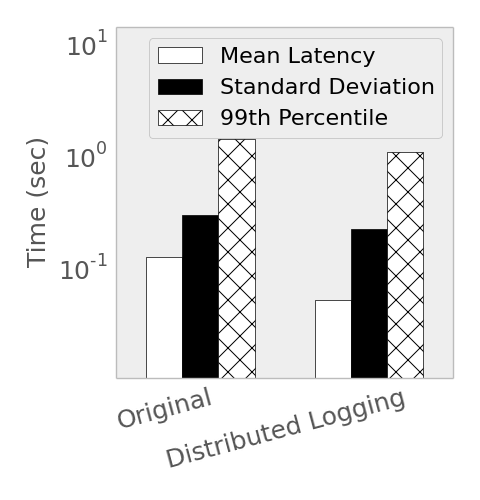}
    \end{subfigure}
    ~
    \centering
    \begin{subfigure}[t]{0.2\textwidth}
        \includegraphics[scale=0.35]{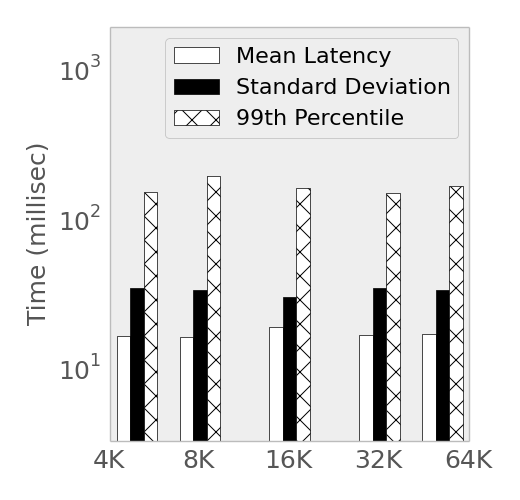}
    \end{subfigure}

    \caption{Effect of distributed logging and redo log block size on Postgres (TPC-C).}
    \label{fig:post_all}
\end{figure}

As discussed in Section~\ref{sec:psql:sol:dist}, we  implement a simple distributed logging scheme for 
Postgres.  Figure \ref{fig:post_all}(left) shows that this technique significantly reduces mean, variance and 99th percentile  
latencies by 58.8\%, 44\% and 23.6\%, respectively.

Another strategy for reducing the variance of redo log flushes in Postgres is to 
	accelerate the I/O operations through tuning an appropriate block size (see Section~\ref{sec:va-tuning}). 
	 In Postgres, redo logs are composed of blocks of the same size, 8 KB by default.
	  Figure \ref{fig:post_all}(right) shows that increasing the block size can reduce variance, but only to a certain extent. 
		This is because a larger block size can reduce the number of write operations per transaction.
		However, when the block size is too larger,  
	 the generated
	   log records only occupy a small portion of a block, while the transaction still has to write a whole block.	  
	  In other words, the disadvantage of writing more data than actually needed eventually outweighs the advantage of fewer writes.

\subsection{Evaluation of VProfiler} 
\label{ssec:vprofiler}
In this section, we evaluate the performance \emph{overhead} of \vprofiler\ in measuring the execution time variance of a function, 
as well as its \emph{efficiency} in narrowing down the search for the main sources of variance.
Note that we have already validated 
\vprofiler's  \emph{effectiveness} in this regard, by showing that our algorithmic and tuning changes, which were informed by  \vprofiler's findings, indeed reduce variance in  MySQL and Postgres.

\begin{figure}[t]
\inv\sinv
    \centering
    \begin{subfigure}[t]{0.3\textwidth}
        \includegraphics[scale=0.2]{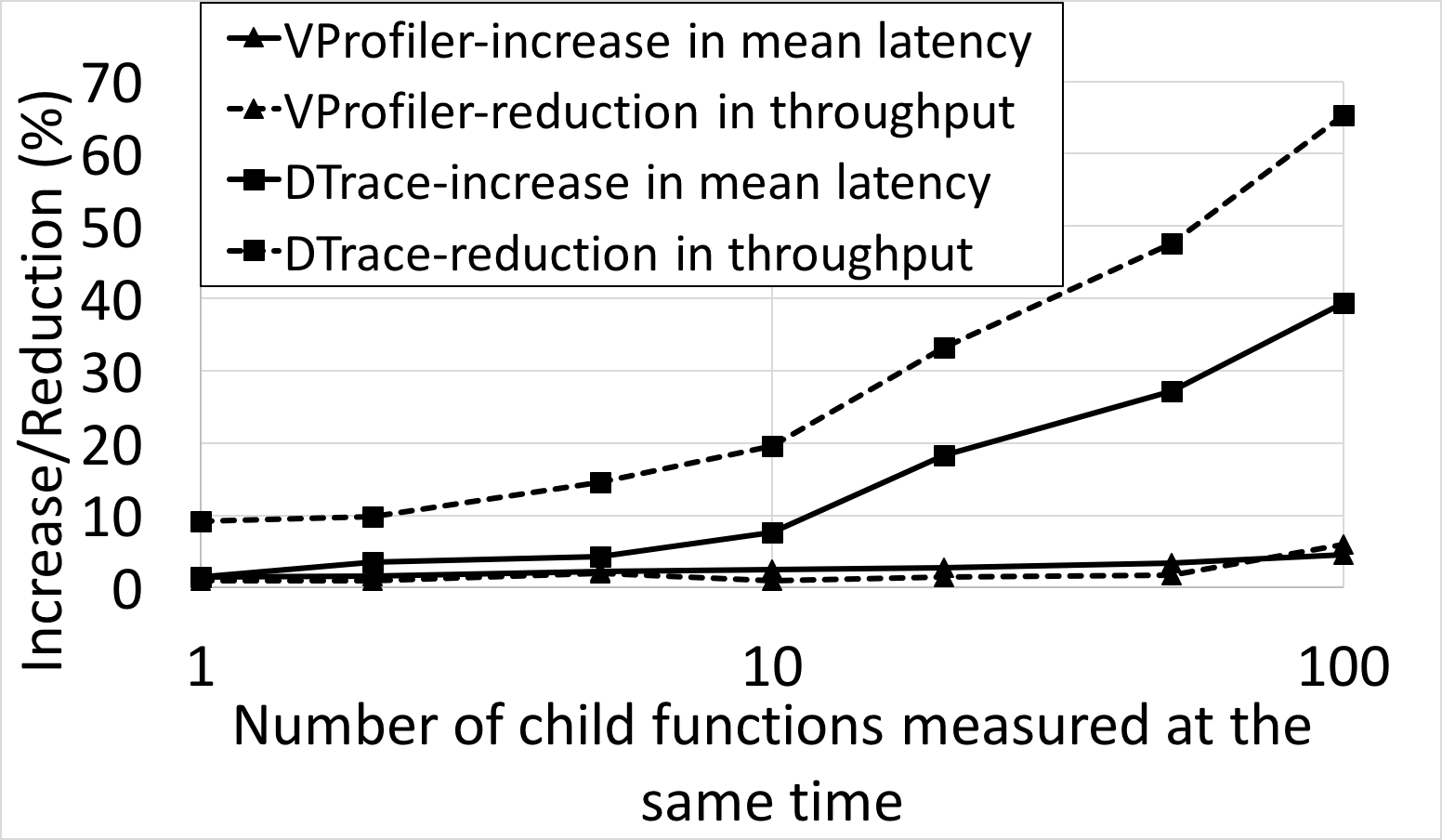}
    \end{subfigure}
    \centering
    \begin{subfigure}[t]{0.15\textwidth}
        \includegraphics[scale=0.2]{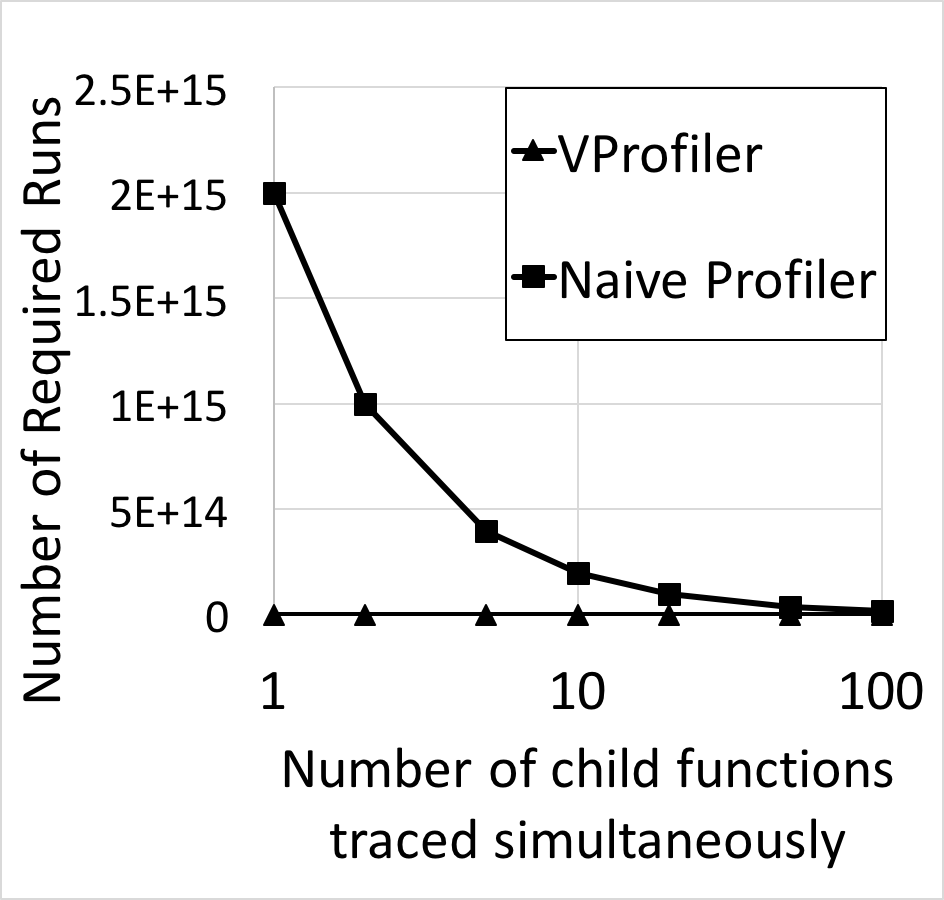}
    \end{subfigure}

\sinv
    \caption{(Left) Profiling overhead of \vprofiler\ vs. DTrace. 
    (Right) Number of runs needed for the profiler to identify the main sources of variance.}
    \inv
    \label{fig:vp_all}
\end{figure}

\ph{VProfiler versus DTrace}
By  instrumentation a DBMS code, \vprofiler\ incurs a performance overhead. 
To quantify this overhead, we vary the number of children functions that need to be instrumented from 1 to 100, and 
	measure both the relative drop of throughput  as well as the relative increase in average latency. 
The results are shown in Figure~\ref{fig:vp_all}(left). 
Here, to provide a baseline, we also report the same types of overhead using DTrace. 

DTrace is a programmable profiler  for troubleshooting arbitrary software. 
\ignore{DTrace is used in an event-listener manner --- users specify the event they want to listen to, along with a piece of code which will be executed when the specified event occurs. Multiple events can be specified at the same time, which are referred to as \textit{probes}. Different types of probes are provided by different \textit{providers}. For example, the \textit{syscall} provider provides probes related to system calls, like the start and exit of the \texttt{open} system call.
DTrace's \textit{pid} provider can dynamically patch running processes with instrumentation code.} 
One can use DTrace to implement a profiler similar to \vprofiler, to measure the execution time of a parent function and its children and then compute variances using Equation~\ref{eq:var-break-down}.

DTrace's key advantage is that, unlike \vprofiler, it does not require the source code for its instrumentation.  
However, this flexibility comes at a cost in the performance of the profiling code.  
We contrast the overhead of DTrace and \vprofiler\ as a function of the number of functions that are instrumented.
 As shown in Figure \ref{fig:vp_all}(left),  the overhead of DTrace (on both latency and throughput) is significantly higher than \vprofiler, and  grows rapidly as the number of 
 traced children increases, whereas the overhead of \vprofiler\ stays below 6\%. 
This is expected as 
DTrace must use heavy-weight mechanisms to inject generalized instrumentation code at run-time, whereas \vprofiler\  inserts minimal profiling code prior to compilation of the MySQL source.


\ph{VProfiler versus a Na{\"i}ve profiler}
Here, we compare \vprofiler\ to a na\"{\i}ve profiler, which is  similar to \vprofiler, except that it breaks down every factor possible instead of only a few important ones. In total, there are $2 \times 10^{15}$ nodes in the static call graph of MySQL, $4.5 \times 10^{14}$ of which are leaves. Since the na\"{\i}ve profiler has to break down every non-leaf node, the number of runs needed is extremely large. The selection strategy of \vprofiler\ greatly reduces the number of runs needed to locate the main sources of variance. 
Figure~\ref{fig:vp_all}(right) confirms this observation.

\subsection{Parameter Tuning for VATS} 
\label{ssec:para-tuning}

\begin{figure*}[t]
\inv\inv
    \centering
    \includegraphics[width=0.9\textwidth]{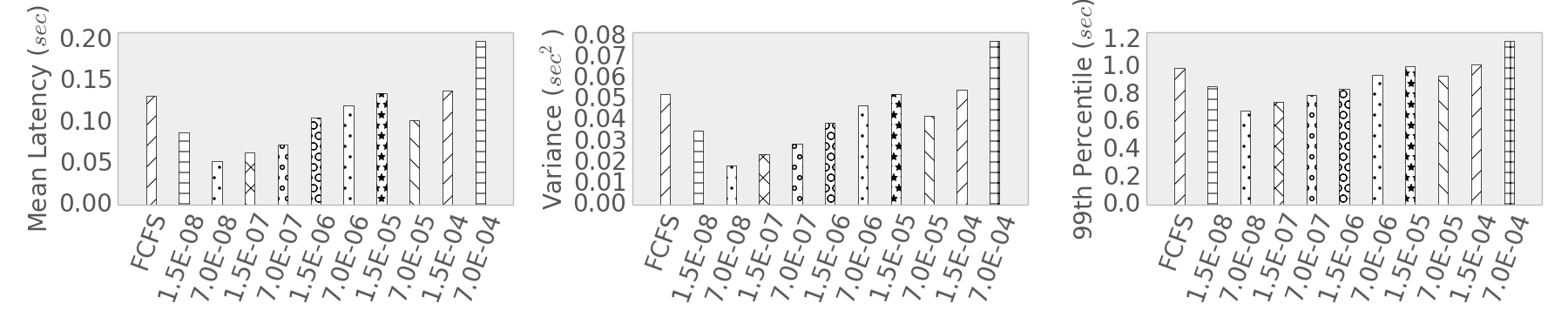}
    \sinv
    \caption{Auto-tuning \vats's threshold value for TPC-C.}
    \inv
    \label{fig:vats-tpcc}
\end{figure*}

Our \vats\ algorithm uses the eldest-thus-far-first scheduling 
 whenever the ratio of the number of wait locks to the total number of locks is greater than some threshold $\theta>0$, and uses FCFS otherwise.
The intuition is that, in the absence of sufficient contention, the choice of the scheduling algorithms is irrelevant and thus, FCFS is preferred due to its lower overhead.
By observing the current latency variance in the system, \vats\ can automatically tune its optimal $\theta$ parameter.
Figure~\ref{fig:vats-tpcc} shows how the overall performance changes with different threshold values. 
As the value of $\theta$ decreases, 
the likelihood of invoking the eldest-thus-far policy increases, which in turn reduces variance. 
However, this trend stops beyond a certain point, when $\theta$ becomes too small (e.g., $\theta\leq 7.0E-8$)
then the amount of contention in the system is not sufficient  to warrant and justify the sorting overhead of the eldest-thus-far scheduling 
(compared to a simple FCFS policy). 




\section{Related Work}
\label{sec:related}
Although it is rarely a focus, performance predictability has been examined in  several broader contexts.

\ph{Query Progress Estimation}
There is a large body of work in progress indicators during the execution of long-running queries \cite{luo:toward, luo:increasing, chaudhuri:can, chaudhuri:estimating}  
and multi-query workloads~\cite{luo:multi}.  
Predicting individual  transaction latencies has been a much harder problem, e.g., 
	only aggregate  resources (e.g., average CPU or disk usage) have been predicted for transactional workloads~\cite{oltp-prediction11,mozafari_pvldb2015_demo,mozafari_sigmod2013}.
Others seek to predict the total runtime of a query workload before it begins execution, by modeling the interactions of a set of queries~\cite{ahmad:interaction}, using machine learning~\cite{ganapathi:predicting, gupta:pqr}, or via sampling and modeling techniques~\cite{duggan:performance}. 
Rather than passive  prediction of performance, we focus on \emph{achieving} predictability though algorithmic and tuning changes of the DBMS.
Also, we work at the drastically finer time scale of transaction latency, where different sources of variance apply. 
 Furthermore, our work focuses on understanding and managing latency variance rather than average performance.

\ph{Real-Time Databases}
Once an active area of research in the 1990's, 
real-time databases (RTDBs)~\cite{aranha:implementation, o:two, abbott:scheduling, huang:experimental, lee:conflict, pang:multiclass, sha:concurrency, kim:approach, kim:supporting, kim:predictability} 
sought real-time performance guarantees by 
(i) requiring each transaction to provide its own deadline,
and (ii) minimizing deadline violations by
restricting themselves to mechanisms that  bounded worst case execution times. 
     In contrast, we study predictability in the context of today's conventional best-effort transaction processing systems,
where  maximizing throughput remains an important optimization goal, and    
optimizations that sacrifice mean latency to obtain hard bounds on execution time may not present an acceptable trade-off.

\ph{Architecting for Predictability}
Some authors have argued for radical DBMS redesign.  
For example, Chaudhuri and Weikum~\cite{chaudhuri:rethinking} argue for ``RISC-like component'' approach to DBMS software design to reduce coupling among sub-systems and make it easier to tune performance.  Florescu and Kossman~\cite{florescu:rethinking} argue that predictability has never been a DBMS design goal and propose a new tiered architecture for web-based applications, where consistency maintenance is moved from the storage layer (i.e., the DBMS) to the application layer. 
Radical redesign of DBMS architecture may have numerous consequences beyond predictability; in this work, we instead seek to understand and mitigate the root causes of unpredictability in existing transaction processing architectures, which may in turn inform future redesign efforts.

\ph{Variance-aware Query Planning}
Instead of a ground-up redesign of database systems, Babcock and Chaudhuri argue for a more practical approach by
modifying the query optimizer to explicitly consider variability in its cost 
formula~\cite{babcock:optimizer}. Their technique projects query performance over a distribution of possible selectivities, and scores query plans based on a weighted mix of their mean and variability of performance. 
Similar to progress indicators, their approach is more appropriate for long-running decision support queries than online transaction processing.  
The sources of latency we target occur at finer time-scales and are not visible at the abstraction level of the query planner.



\ph{Variance-Aware Job Scheduling}
Beyond our database context, theoretical literature has examined the problem of scheduling general tasks to minimize completion time variance (CTV) and waiting time variance (WTV).  In these problem formulations, there is a queue of jobs with \emph{known processing times} waiting to be scheduled, K jobs at a time, and the goal is to find a scheduling order that 
minimizes  the variance of the completion or waiting times of  the jobs.
 While CTV and WTV problems are both NP-complete~\cite{kubiak:completion,merten:variance}, some properties of optimal orderings are nevertheless known. 
Most notably, it is proven that an optimal schedule has a so-called ``V-shape property''~\cite{bector:v-shape,krieger:v-shape,cai:v-shape},
which means that the job with the greatest processing time must be scheduled first, followed by a subset of 
other jobs in their decreasing order of processing times, followed by the remaining jobs in their increasing order of processing times.
However, the V-shape property, only helps in determining which job to schedule first. 
For scheduling the remaining jobs, 
there are several heuristics~\cite{eilon:heuristic, kanet:heuristic, vani:heuristic, chen:heuristic, ye:heuristic}, 
dynamic programming solutions~\cite{kubiak:dynamic, de:dynamic}, and a polynomial-time approximation~\cite{kubiak:approximate}. 

These techniques assume an \emph{offline} setting, and thus do not apply to our transaction scheduling problem, since 
the processing time and the arrival time of transactions are unknown \emph{a priori}.
In other words, transaction scheduling is an \emph{online} problem, where the system does not know which locks 
will be requested next and how long they will be held once granted. 
 However, our proposed variance-aware transaction scheduling (\vats),
 which uses the eldest-transaction-first lock transfer policy, is 
 in fact inspired by the V-shape-based heuristics to the CTV problem~\cite{vani:heuristic}.


\section{Conclusion}
\label{sec:conclusion}

We presented a novel profiler, called \vprofiler, for automatically identifying the major sources of latency variance in a 
transactional  database. By breaking down the variance of latency into variances and covariances of functions in the source code of the software, \vprofiler\ makes it possible to calculate the contribution of each function to the overall variance.
Using our tool, we analyzed the codebases of two popular DBMSs, leading us to both generic and DBMS-specific 
	solutions for reducing latency variance. In particular, we introduced a new scheduling algorithm that 
	is proven to minimize $L_p$ norm (and reduce mean and tail latencies), a new buffer page replacement, and 
	variance-aware configurations for tuning. All these techniques greatly reduce the variance of transaction latency.
	
	\inv\inv

\let\oldbibliography\thebibliography
\renewcommand{\thebibliography}[1]{\oldbibliography{#1}
\setlength{\itemsep}{0pt}} 

{\small
\bibliographystyle{abbrv}
\bibliography{predictability,mozafari,prediction,robust}
}
\end{document}